\documentclass{els-mrw-arxiv} 

\usepackage[style=review, natbib=true]{biblatex}

\usepackage{newtxtext,newtxmath,helvet}

\usepackage{microtype}
\usepackage{enumitem}
\usepackage{hyperref}
\hypersetup{colorlinks, linkcolor=Tomato, urlcolor=MediumOrchid, citecolor=CornflowerBlue}

\def\equationautorefname~#1\null{Equation~(#1)\null}
 
\newcommand\re{R_\mathrm{e}}
\newcommand\se{\sigma_\mathrm{e}}
\newcommand\lam{\lambda_{\re}}
\newcommand\eps{\varepsilon_\mathrm{e}}
\newcommand\lameps{(\lam, \eps)}
\newcommand\msun{M_\odot}
\newcommand\mcrit{M_*^\mathrm{crit}}
\newcommand\scrit{\mathrm{\sigma_e^{crit}}}
\newcommand\ser{S\'ersic}
\newcommand\kms{\text{ km s}^{-1}}

\begin{document} 
 
\chapter[Early-Type Galaxies: Elliptical and S0 Galaxies, or Fast and Slow Rotators]{Early-Type Galaxies:\\ Elliptical and S0 Galaxies, or Fast and Slow Rotators} 

\author[1]{Michele Cappellari}

\address[1]{\orgname{Sub-Department of Astrophysics}, \orgdiv{Department of Physics}, \orgaddress{University of Oxford, Denys Wilkinson Building, Keble Road, Oxford, OX1 3RH, UK}}

\maketitle

\begin{glossary}[Glossary]

    \term{Elliptical galaxies (E)} are galaxies with smooth, featureless light distributions and elliptical shapes, lacking spiral arms. They are denoted by `E' followed by a number indicating their ellipticity.

    \term{S0 galaxies (lenticular galaxies)} have a central bulge and a disk structure but lack the spiral arms characteristic of spiral galaxies.

    \term{Early-type galaxies (ETGs)} comprise elliptical (E) and lenticular (S0) galaxies, or more meaningfully, are classified as fast or slow rotators based on their kinematic properties.

    \term{Fast-rotator ETGs} are early-type galaxies with significant specific angular momentum, featuring stellar disks evident in their kinematics at any inclination. Extended disks appear as S0 galaxies when edge-on, while centrally concentrated disks with dominant spheroids are classified as disky ellipticals when edge-on. At low inclinations, fast rotators may be misclassified as ellipticals.

    \term{Slow-rotator ETGs} are early-type galaxies with low specific angular momentum, typically spheroidal and lacking stellar disks. They are generally found above a critical stellar mass of $\mcrit\approx2\times10^{11}\,M_\odot$, or more accurately when $\lg(\re/\text{kpc}) \gtrsim 12.4 - \lg(M_*/M_\odot)$. Slow rotators are classified as ellipticals regardless of orientation.

    \term{Supermassive black hole (SMBH)} refers to black holes with masses ranging from $10^6-10^{10}\msun$, typically found at the centers of galaxies. Influences galaxy evolution through feedback mechanisms that regulate star formation.

    \term{Kinematically-decoupled stellar core (KDC)} is a central region in a galaxy where the stars rotate in a different direction or at a different speed compared to the outer parts of the galaxy, often indicating past merger events

    \term{Integral-field Spectroscopy (IFS)} is an observational technique that captures spatially-resolved spectral data across a galaxy, enabling two-dimensional mapping of kinematic and stellar population properties.

    \term{Line-of-sight velocity distribution (LOSVD)} describes the range of velocities of stars or gas along the line of sight in a galaxy, providing insights into the galaxy's kinematics and mass distribution.

    \term{Stellar initial mass function (IMF)} is a function that describes the initial distribution of masses for a population of stars at the time of their formation, influencing the evolution of galaxies and the interstellar medium.

    \term{Half-light or effective radius ($\re$)} is the radius within which half of a galaxy's total light is emitted. It is calculated as the radius of a circle with the same area as the galaxy's half-light isophote.

    \term{Light-weighted second velocity moment ($\se$)} is a kinematic parameter defined as $\se^2 \equiv \langle V^2 + \sigma^2 \rangle$, luminosity-weighted within the effective radius ($\re$). Although often referred to as `effective velocity dispersion,' $\se$ combines contributions from both ordered stellar rotation ($V$) and random motions ($\sigma$).

    \term{Core density ($\Sigma_1$)} is the stellar mass density within an aperture of radius $R=1$ kpc, providing a measure of the central concentration of stars in a galaxy, which is related to its bulge mass fraction.

    \term{Specific angular momentum ($\lambda_R$)} is a dimensionless parameter $0\leq\lambda_R\leq1$ that qualitatively measures the stellar angular momentum per unit mass in a galaxy. This parameter is typically calculated within the half-light isophote, $\re$, and denoted as $\lam$. It is defined by the equation: $\lambda_R = \langle R|V| \rangle/\langle R\sqrt{V^2 + \sigma^2} \rangle$, where \( R \) represents the radius, \( V \) is the rotational velocity, and \( \sigma \) is the velocity dispersion.

    \term{Fundamental Plane} of galaxies is a relation connecting galaxies' half-light radius ($R_R$), total luminosity ($L$), and second velocity moment ($\se$). Galaxies occupy a plane in the space defined by ($\lg \re$, $\lg L$, $\lg \se$) due to virial equilibrium and gradual variations in mass-to-light ratio ($M/L$) with $\se$.

    \term{Mass Plane} of galaxies is a relation linking a galaxy's mass ($M$), size ($\re$), and second velocity moment ($\se$). Galaxies lie on a tight plane in the space defined by ($\lg \re$, $\lg M$, $\lg \se$), primarily due to virial equilibrium.

    \term{Virial mass estimator} is a method to estimate the mass ($M\propto\re\se^2$) or more accurately the mass-to-light ratio ($M/L$) within a given region of a gravitationally bound system using the virial theorem, relating the system's total kinetic energy to its gravitational potential energy.

    \term{Galaxy environment} refers to local conditions affecting galactic evolution, ranging from isolated fields to dense clusters and cosmic filaments.

    \term{Stellar metallicity} is the abundance of elements heavier than hydrogen and helium, indicating the galaxy's chemical composition and star formation history. Higher metallicity suggests multiple generations of star formation, enriching the interstellar medium with metals.

    \term{Stellar $\alpha$ enhancement} refers to a higher ratio of $\alpha$ elements (like oxygen and magnesium) to iron, indicating a rapid, early star formation period. $\alpha$ elements are produced by Type II supernovae from massive stars (lifetime $<5\times10^7$ yr), while iron is produced by Type Ia supernovae over longer timescales.

    \term{Galaxy quenching} is the cessation of star formation in a galaxy due to environmental effects, internal feedback mechanisms, or gas depletion.

    \term{Galaxy clusters} are vast structures containing hundreds to thousands of galaxies bound by gravity, embedded in dark matter and hot, X-ray-emitting gas. They are crucial for studying large-scale structure formation and galaxy evolution.

\end{glossary}

\newpage

\begin{abstract}
    Early-type galaxies (ETGs) show a bimodal distribution in key structural properties like stellar specific angular momentum, kinematic morphology, and nuclear surface brightness profiles. Slow rotator ETGs, mostly found in the densest regions of galaxy clusters, become common when the stellar mass exceeds a critical value of around $\mcrit\approx2\times 10^{11}\,M_\odot$, or more precisely when $\lg(\re/\mathrm{kpc}) \gtrsim 12.4 - \lg(M_*/M_\odot)$. These galaxies have low specific angular momentum, spheroidal shapes, and stellar populations that are old, metal-rich, and $\alpha$-enhanced. In contrast, fast rotator ETGs form a continuous sequence of properties with spiral galaxies. In these galaxies, the age, metallicity, and $\alpha$-enhancement of the stellar population correlate best with the effective stellar velocity dispersion $\se \propto \sqrt{M_*/\re}$ (i.e., properties are similar for $\re\propto M_*$), or with proxies approximating their bulge mass fraction. This sequence spans from star-forming spiral disks to quenched, passive, spheroid-dominated fast rotator ETGs. Notably, at a fixed $\se$, younger galaxies show lower metallicity. The structural differences and environmental distributions of ETGs suggest two distinct formation pathways: slow rotators undergo early intense star formation followed by rapid quenching via their dark halos and supermassive black holes, and later evolve through dry mergers during hierarchical cluster assembly; fast rotators, on the other hand, develop more gradually through gas accretion and minor mergers, becoming quenched by internal feedback above a characteristic $\lg(\scrit/\kms)\gtrsim2.3$ (in the local Universe) or due to environmental effects.
\end{abstract}

\begin{BoxTypeA}[boxlabel]{Key Points}  
    \renewcommand{\labelitemii}{+}  
    \begin{itemize}  
        \item Early-type galaxies (ETGs) exhibit a bimodal distribution (two distinct peaks) in structural properties:  
        \begin{itemize}  
            \item Stellar specific angular momentum  
            \item Kinematic morphology  
            \item Nuclear surface brightness profiles  
        \end{itemize}  
        \item The traditional E/S0 morphological classification fails to reflect this bimodality due to inclination-related observational biases
        \item A rotation-based classification (fast vs. slow rotators) provides a more physically meaningful framework:  
        \begin{itemize}  
            \item Slow rotators:  
            \begin{itemize}  
                \item Primarily reside in dense regions of galaxy groups/clusters  
                \item Exhibit strong mass dependence: dominate above \( \mcrit \approx 2 \times 10^{11} \msun \)  
                \item Sharper separation including half-light radii: \( \lg(\re/\text{kpc}) \gtrsim 12.4 - \lg (M_*/M_\odot) \)  
                \item Features: Low specific angular momentum, spheroidal morphology, old/metal-rich/\(\alpha\)-enhanced stellar populations  
            \end{itemize}  
            \item Fast rotators:  
            \begin{itemize}  
                \item Host stellar disks and form a continuous sequence of physical properties with spiral galaxies  
                \item Stellar population properties (age, metallicity, \(\alpha\) enhancement) best correlate with \( \se \propto \sqrt{M_*/\re} \)  
                \item Implication: Population properties remain constant along \( M_* \propto \re \) lines in the \((M_*, \re)\) plane  
                \item \( \se \) empirically traces bulge mass fraction or central stellar density at fixed galaxy mass  
                \item Sequence ranges from star-forming disks (bulge-less) to quenched spheroid-dominated fast rotators  
            \end{itemize}  
        \end{itemize}  
        \item Formation pathways differ markedly between the two classes:  
        \begin{itemize}  
            \item Slow rotators:  
            \begin{itemize}  
                \item Early rapid star formation followed by quenching via dark halo/SMBH feedback  
                \item Growth via dry mergers during hierarchical assembly of host groups/clusters  
            \end{itemize}  
            \item Fast rotators:  
            \begin{itemize}  
                \item Gradual evolution through gas accretion and minor mergers  
                \item Quenching triggered by internal feedback, for $\lg(\scrit/\kms)\gtrsim2.3$, or environmental processes  
            \end{itemize}  
        \end{itemize}  
    \end{itemize}  
\end{BoxTypeA}

\section{Introduction}
\label{sec:intro} 
  
``What are galaxies? No one knew before 1900. Very few people knew in 1920. All astronomers knew after 1924'' With this sentence, Allan Sandage opens his book \textit{The Hubble Atlas of Galaxies}. The year 1924 marked a pivotal moment in astronomy with Edwin Hubble's groundbreaking work, which convincingly demonstrated that galaxies are extragalactic objects. Today, a century later, we understand that galaxies are fundamental building blocks of the Universe. These cosmic structures, containing billions of stars, have sizes and 
masses comparable to the Milky Way galaxy.  
 
The discovery that galaxies are `island Universes' initiated their systematic study. As in many scientific fields, the first step involved classifying them to uncover regularities in their properties. Extending earlier, more rudimentary approaches \citep[see a historical review by][]{Sandage2005}, Edwin Hubble proposed a well-defined scheme for visually classifying galaxies based on optical images \citep{Hubble1926}. Two major classes emerged: (i) Early-Type Galaxies (ETGs), characterized by their smooth, elliptical or lenticular shapes without spiral arms, and (ii) Spiral galaxies, defined by the presence of spiral arms traced by patchy dust and clumpy star-forming regions. In this chapter, focusing on ETGs, I consistently use Hubble's morphological classification, which remains universally adopted, to distinguish them from spiral galaxies.

ETGs dominate the high-mass end of the galaxy mass distribution. In our current paradigm of galaxy formation, galaxies and their central supermassive black holes (SMBHs) grow hierarchically, starting from smaller building blocks and increasing in mass over time \citep{White1978, Blumenthal1984}. This suggests that ETGs represent the end-point of galaxy and SMBH evolution. As such, they encode the full history of galaxy assembly and are ideal for studying galaxy formation and evolution. By examining ETG properties such as their stellar populations, kinematics, and chemical abundances, astrophysicists can reconstruct their formation and evolutionary histories. 

Moreover, ETGs offer a unique laboratory for investigating how galaxies evolve through interactions with their surrounding environment. Processes like galaxy mergers, harassment, and ram pressure stripping can significantly impact ETG properties, providing clues about the environmental factors shaping their evolution.

This chapter provides a comprehensive overview of ETGs, exploring their defining characteristics, classification schemes, and fundamental properties. I will discuss their formation and evolution, considering both theoretical models and observational evidence. The impact of environmental factors on ETGs will also be discussed, highlighting the complex interplay between galaxies and their surroundings.
 
\section{Photometric Properties}

\subsection{Morphological Classification from Images}\label{sec:morphology}

\begin{figure*} 
    \centering
    \includegraphics[width=\textwidth]{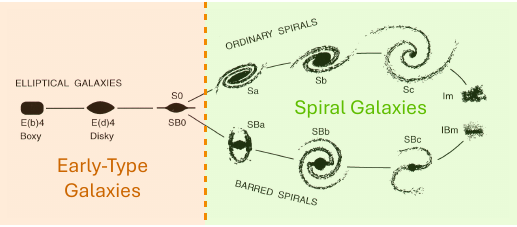} 
    \caption{Hubble's tuning-fork diagram for galaxy classification, adapted from \citet[pg.~45]{Hubble1936}. The diagram features elliptical galaxies (E) along the handle, which then splits into two prongs representing spiral galaxies. The top prong depicts normal spirals (S), ranging from tightly wound, bulge-dominated types (Sa) to loosely wound, bulge-less types (Sc). The bottom prong shows barred spirals (SB), characterized by a bar-shaped structure and similar subclasses. Lenticular galaxies (S0), located at the fork's split, bridge the gap between elliptical and spiral galaxies, featuring a central bulge and disk but lacking prominent spiral arms. This version of the diagram was modified by \citet{Kormendy1996} to include boxy E(b) and disky E(d) elliptical galaxies, as well as irregular galaxies Im. I added annotations to differentiate early-type galaxies (E and S0), the focus of this chapter, from spiral galaxies. \label{fig:tuning_fork}}
\end{figure*} 

The term `early' in ETGs was inspired to Hubble by the evolutionary sequence proposed by \citet{Jeans1928}, who also presented a Y-shaped diagram to visually arrange morphological classes, a precursor to Hubble's tuning-fork. In the first published version of his classification, \citet{Hubble1926} referred to all ETGs as `elliptical' galaxies. Later, in Chapter II of the book where he presented his classification tuning fork \citep{Hubble1936}, he introduced S0 galaxies as a ``hypothetical class'' intermediate between elliptical and spiral galaxies.

Hubble classified elliptical galaxies (Es) based on their apparent axial ratio. Ellipticity is defined as $\varepsilon=1 - b/a$, where $a$ and $b$ are the major and minor axes of the galaxy isophotes, respectively. Hubble defined the position in the E sequence by estimating the ellipticity to one decimal point, omitting the point (e.g., an elliptical with $\varepsilon\approx0.32$ is classified as E3). Hubble later recognized that variations in elliptical galaxies could not be described by ellipticity alone, and that Es needed to be separated into (i) ellipticals (E), with elliptical isophotes, and (ii) S0 galaxies, with lenticular shapes flatter than E7 but no evidence of spiral arms. This revised Hubble classification, still universally adopted today, was published after his death by his collaborator Allan Sandage \citep{Sandage1961}. \autoref{fig:tuning_fork} presents Hubble's tuning fork, with the ETGs part of the diagram modified by \citet{Kormendy1996} to account for key differences in the elliptical class.

Spiral galaxies are characterized by the presence of a stellar disk and spiral arms. At the center of the disk, they contain a stellar spheroid or `bulge' which somewhat resembles a very flattened E galaxy. Hubble classified spiral galaxies as Sa, Sb, and Sc according to the prominence of the bulge, with Sa being the most bulge-dominated. He also noted that more bulge-dominated spirals have more closely coiled spiral arms and used this as an additional criterion in his classification.

Later, \citet{vandenBergh1976} noted that S0 galaxies can span the full range of bulge fractions as spiral galaxies. He proposed a revision of Hubble's classification by adding a small letter to S0 galaxies, like S0a, S0b, and S0c, to qualitatively indicate their bulge fraction. He suggested a trident to replace the tuning fork, with the S0s in one of the arms like the spirals, to highlight the parallelism between spiral and S0 galaxies. A problem with this classification is that S0 galaxies are not easy to distinguish from E galaxies unless seen close to edge-on, because bulges become hard to recognize from the stellar disk. Bulge fractions of S0s are even more difficult to visually assess at low inclination ($i=90$ being edge-on).

Although the distinction between early-type and spiral galaxies is based on visual images alone, the two types differ by various other characteristics, indicating that the classification is physically meaningful. ETGs appear smooth because they have lower amounts of gas and star formation than spiral galaxies. Therefore, their stellar population is generally older and they have redder colors. However, a separation based on color or ages is not fully equivalent to one based on morphology and cannot be used interchangeably \citep{Strateva2001}.

\subsection{Global Surface Brightness Profiles}\label{sec:sersic_profiles}

\begin{figure*}
    \centering
    \includegraphics[width=.49\textwidth]{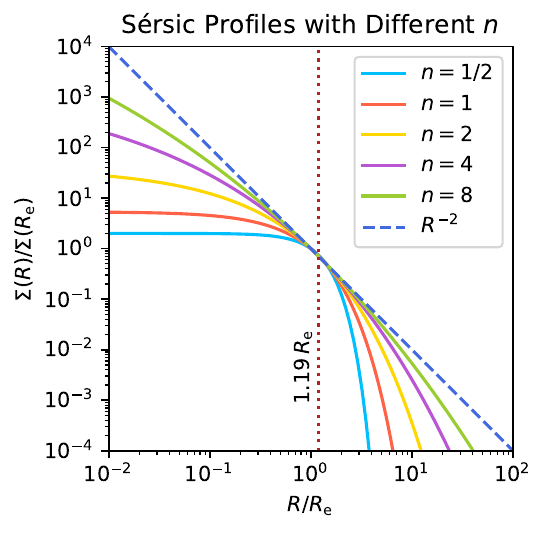}
    \includegraphics[width=.49\textwidth]{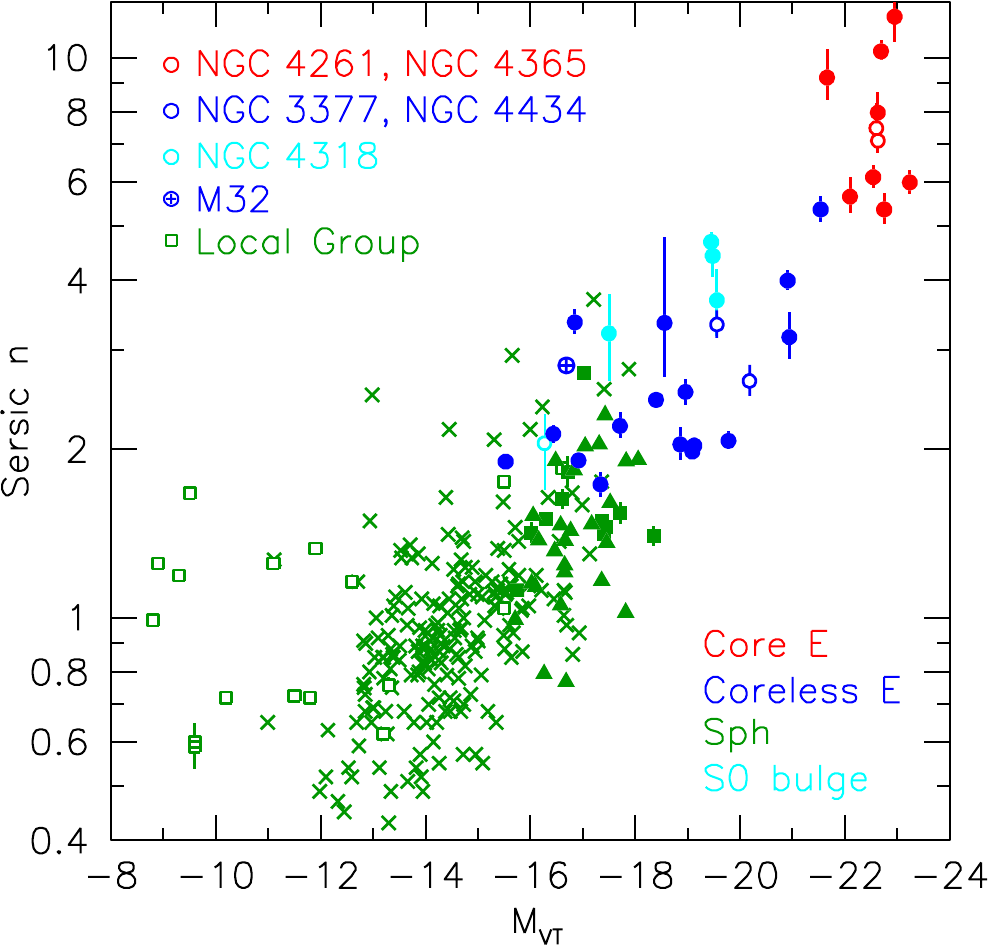}
    \caption{
        Left panel: \ser\ profiles of \autoref{eq:sersic} for different logarithmically-spaced values of the \ser\ index $n$. All profiles have the same logarithmic slope $\gamma=-2$ at a radius $R\approx1.19\re$, indicated by the vertical dotted line with 1\% accuracy, where $\re$ is the half-light radius. The dashed line shows the asymptotic profile $\Sigma\propto R^{-2}$ for $n\rightarrow\infty$.
        Right panel: Correlation between \ser\ index $n$ and visual magnitude $M_{VT}$. Red, blue, green, and cyan points represent core ellipticals (Es), non-core ellipticals, spheroidal galaxies, and S0 bulges, respectively. Green triangles, crosses, and open squares indicate spheroidals \citep[fig.~33]{Kormendy2009}.
        \label{fig:sersic}}
    \end{figure*}

Early quantitative measurements of the surface brightness profiles $\Sigma(R)$ of ETGs, obtained from digitized photographic plates, initially suggested that they could be universally described by an $R^{1/4}$ function when expressed in magnitudes $\mu=-2.5\lg (\Sigma/\Sigma_0)$ \citep{deVaucouleurs1948}. However, later observations, leveraging the development of CCD detectors, revealed that these profiles were not universal. Instead, they required a more general function proposed by \citet{Sersic1968}:
\begin{equation}
    \Sigma(R)=\Sigma_\mathrm{e} \exp\left\{-b_n \left[ \left(\frac{R}{\re}\right)^{1/n} - 1\right]\right\},
    \label{eq:sersic}
\end{equation}
This function reduces to the de Vaucouleurs profile when $n=4$. Here, $\re$ represents the radius enclosing half of the total galaxy luminosity integrated from the analytic profile, $\Sigma_\mathrm{e}\equiv\Sigma(\re)$, and $b_n = Q^{-1}(2n,1/2)$ is a normalization factor designed to satisfy the $\re$ definition, with $Q^{-1}$ the inverse of the regularized incomplete gamma function (\citealt{Olver2010nist}, \href{https://dlmf.nist.gov/8.2#E4}{eq.~8.2.4}), which is available in most popular programming languages\footnote{E.g., in Python \href{https://docs.scipy.org/doc/scipy/reference/generated/scipy.special.gammainccinv.html}{scipy.special.gammainccinv}, or in Mathematica \href{https://reference.wolfram.com/language/ref/InverseGammaRegularized.html}{InverseGammaRegularized}.} (Kai Zhu et al.\ in preparation).

Qualitatively, as $n$ increases, the galaxy light at radii around $\re$ is redistributed to both smaller and larger radii (see \autoref{fig:sersic}). A large $n$ indicates a galaxy with both (i) a more concentrated light distribution in the central regions and (ii) a more extended stellar halo at larger radii. In the limit as $n\rightarrow\infty$, the \ser\ profile tends to a power law $\Sigma(R)\propto R^{-2}$. As $n$ increases, a larger fraction of the total luminosity of the \ser\ profile is distributed to larger radii, where the surface brightness can fall below observable levels. This makes the empirical determination of $\re$ less accurate for galaxies with large $n$. \autoref{fig:sersic} also shows that the variation of the \ser\ profiles is nearly uniform when $n$ is sampled logarithmically. Therefore, correlations involving the \ser\ index are better studied using $\lg n$ instead of $n$, to avoid the `saturation' of profile variations at large $n$.

As shown in \autoref{fig:sersic}, the half-light radius $\re$ is close to the radius $R_2$ at which the \ser\ profiles have the logarithmic slope $\gamma=-2$ of the asymptotic profile. More precisely, $R_2/\re\approx1.19$ with better than 1\% accuracy for all realistic values of $n$. For a Gaussian ($n=1/2$ in \autoref{eq:sersic}), $R_2/\re=1/\sqrt{\ln(2)}\approx1.201$, and this ratio monotonically decreases with a minimum value $\lim_{n\rightarrow\infty}R_2/\re=e^{1/6}\approx1.181$.

The importance of \ser\ profiles lies not just in providing better fits to galaxy surface brightness. More importantly, $n$ correlates with galaxy properties, with brighter and larger galaxies having larger $n$ \citep{Caon1993}. \autoref{fig:sersic} illustrates the correlation between galaxy luminosity and $n$ as derived by \citet{Kormendy2009}. This trend indicates that the \ser\ index contains information on galaxy evolution, as I will discuss later.

A single \ser\ profile generally does not describe S0 galaxies well, as they often consist of a bulge and a disk component like spiral galaxies. In these cases, the disks are generally described by exponential profiles ($n=1$ in \autoref{eq:sersic}) like those of spirals \citep{Freeman1970}, and only the bulge has a free \ser\ index $n$.

\subsection{Nuclear Surface Brightness Profiles}
\label{sec:inner_slopes}

\begin{figure*}
    \centering
    \includegraphics[width=\textwidth]{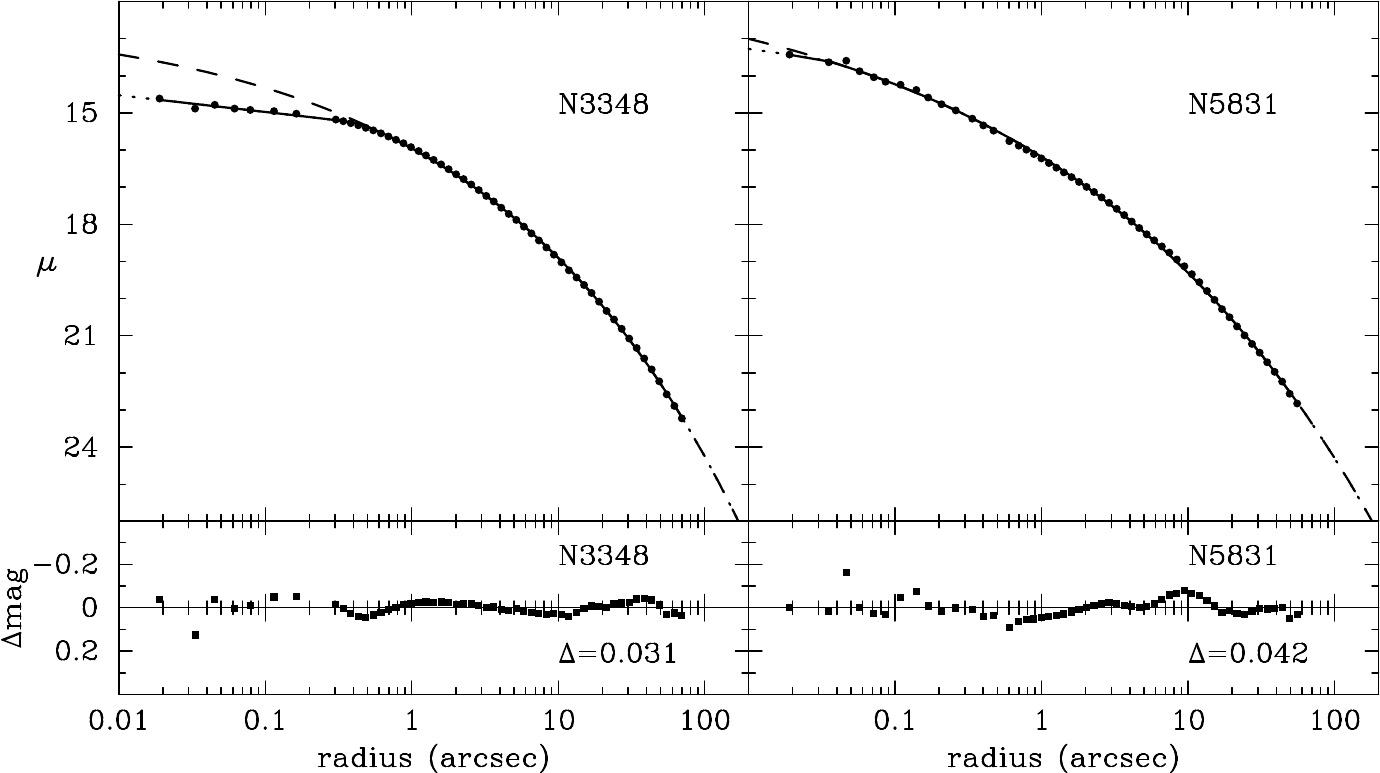}
    \caption{HST major-axis surface brightness profiles of elliptical galaxies. The solid lines represent fits using \autoref{eq:core_sersic}, with dotted lines showing inner and outer extrapolations, and dashed lines indicating inward extrapolations of the \ser-like component beyond the break radius. For NGC 5831, the best fit is nearly a pure \ser\ model. In contrast, NGC 3348 (classified as a `core' galaxy) exhibits a distinct inner break. The root-mean-square (rms) scatter $\Delta \mathrm{mag}$ for each fit is provided below the corresponding panel \citep[fig.~10]{Graham2003core}. \label{fig:core_sersic}}
\end{figure*}

With the launch of the Hubble Space Telescope (HST) in 1990, it became possible to measure galaxy surface brightness profiles at radii smaller than the atmospheric seeing limit, which typically blurs point-like stars to about 1 arcsecond in size (full-width at half-maximum, FWHM). Observations revealed that the surface brightness profiles of ETGs continue to increase as a power law $\Sigma(R)\propto R^{-\gamma}$ down to the HST's spatial resolution limit. However, some galaxies exhibited a nuclear break with a sharp change in the logarithmic slope $\gamma$ inside the break radius \citep{Ferrarese1994, Lauer1995}.

The nuclear surface brightness profiles can be described using either double power-laws \citep{Lauer1995} or the Core-\ser\ parametrization \citep{Graham2003core, Trujillo2004}:
\begin{equation}\label{eq:core_sersic}
    \Sigma(R) = \Sigma' 
    \left[1 + \left(\frac{R_\mathrm{b}}{R}\right)^\alpha\right]^{\gamma/\alpha}
    \exp\left[-b_n \left(\frac{R^\alpha + R_\mathrm{b}^\alpha}{\re^\alpha}\right)^{1/(\alpha\, n)}\right].
\end{equation}
Outside the inner break at radius $R=R_b$, this is a \citet{Sersic1968} profile with a projected half-light radius $\re$, but it gradually transitions to a power-law surface brightness $\Sigma(R)\propto R^{-\gamma}$ at smaller radii $R\ll R_\mathrm{b}$. The exponent $\alpha$ controls the sharpness of the break, while $\Sigma'$ is the overall normalization. \autoref{fig:core_sersic} shows an application of this profile to a real galaxy and illustrates the difference between core/deficit galaxies and pure \ser\ galaxies.

The behavior of the nuclear surface brightness is strongly related to the galaxy's total luminosity. Luminous galaxies with absolute total $V$-band magnitudes $M_V\lesssim-22$ have a core, characterized by a sharp break from a steep outer profile to a much shallower inner profile with $\gamma\lesssim 0.3$. The core can also be described as a light \emph{deficit} relative to a \ser\ profile and quantified using the profile in \autoref{eq:core_sersic}. At lower luminosities, galaxies have inner surface brightness profiles described by a \ser\ profile, with some overlap in the transition region \citep[e.g.,][]{Faber1997}. The fact that bright ETGs posses both a large \ser\ index and a core/deficit in their nuclei is illustrated in \autoref{fig:sersic}.

\subsection{Isophotal Shapes}

With the advent of CCD detectors in the 1980s, it became possible to analyze the detailed shape of the isophotes of elliptical galaxies. It was discovered that some of these galaxies were not well approximated by pure ellipses. Instead, they had more disky, or almond-shaped, isophotes. This was interpreted as indicating the presence of stellar disks, which were not as prominent as the disks of S0 galaxies and could only be detected as deviations of their isophotes from ellipses \citep{Bender1988}.

Bright galaxies, unlike fainter ones, lacked disky isophotes, contained cores in their inner profiles, and exhibited slow rotation (\autoref{sec:kinematics}). This was deemed significant enough to propose a revision of Hubble's tuning-fork diagram. \citet{Kormendy1996} suggested that there was a continuous sequence from S0 galaxies to disky ellipticals, while brighter ETGs with cores constituted a distinct class of objects. They proposed using isophotal shape in galaxy classification, and their revised tuning-fork diagram is shown in \autoref{fig:tuning_fork}.

There was a crucial practical issue with the proposed classification: the presence of stellar disks in ellipticals could only be revealed by the shape of their isophotes when galaxies were close to edge-on ($i\approx90^\circ$). Disky ellipticals would display elliptical isophotes in projection at lower inclinations. This limitation prevented the application of the proposed scheme on a galaxy-by-galaxy basis. This inclination dependence was the same issue that hindered the practical application of the S0a--S0c scheme proposed by \citet{vandenBergh1976}. As I will show in the next section, a solution to the inclination dependence of photometric classification schemes is provided by stellar kinematics obtained with integral-field spectroscopic observations.

\section{Kinematics and Dynamics}\label{sec:kinematics}

\subsection{Kinematic Morphology}

\begin{figure*} 
    \centering
    \includegraphics[width=\textwidth]{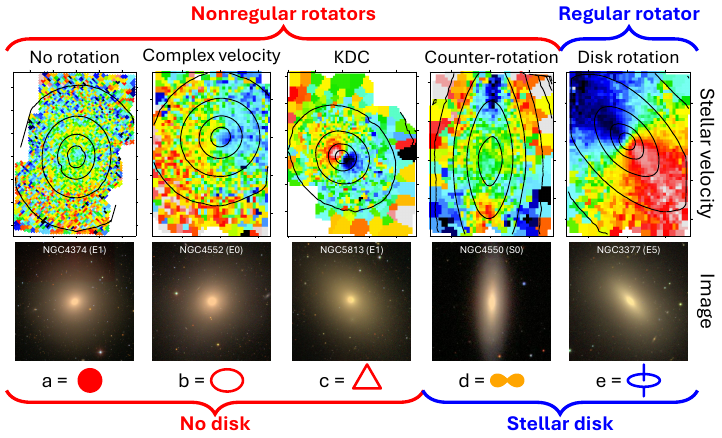}    
    \caption{Morphological classifications based on stellar kinematics. Early-type galaxies are divided into five classes as defined by \citet{Krajnovic2011}: (a) no detectable rotation, (b) nonregular rotation, (c) kinematically distinct cores (KDCs), (d) counter-rotating disks, and (e) regular disk-like rotation. The kinematic data, obtained from \citet{Emsellem2004}, were spatially binned using the Voronoi tessellation method of \citet{Cappellari2003}. Tick marks on the images are spaced at 10 arcsecond intervals. The bottom row displays corresponding SDSS images, with Hubble's morphological classifications indicated in parentheses next to the galaxy names. Classes (d) and (e) are physically related, as both feature stellar disks and are nearly axisymmetric, while the other classes show no evidence of stellar disks. Class (e) is labeled as regular rotators, whereas the remaining classes are classified as nonregular rotators. The symbols below the images represent the different morphological classes and are also used in \autoref{fig:kin_mis} and \autoref{fig:lam_eps}. \label{fig:kinematic_morphology}}
\end{figure*}

The beginning of this century saw the emergence of integral-field spectroscopic (IFS) observations of galaxies. This allowed astrophysicists to obtain spectra over a contiguous two-dimensional region on the sky, fully covering the central regions of the galaxies, typically out to about $1\re$, where the surface brightness is sufficiently large.

Similar to photometry, the first step after observing a statistically significant sample of galaxies with IFS was to visually classify the kinematic maps. All the maps of ETGs could be divided into five main classes introduced by \citet[table~3]{Krajnovic2011} and illustrated in \autoref{fig:kinematic_morphology}. The classification relies on the visual appearance of the maps, not the amplitude of the rotation.

\begin{description}
    \item[(a) No rotation:] Galaxies showing barely detectable rotation at the typical accuracy level ($\sim5 \kms$) of good stellar kinematic measurements. The kinematic maps of these galaxies appear noisy when scaled to the maximum/minimum range of values.
    
    \item[(b) Complex rotation:] Galaxies presenting some clear, organized, and point-symmetric rotation with twists in the stellar kinematics, without a clearly defined symmetry axis in the kinematic map.
    
    \item[(c) Kinematically-decoupled core:] Galaxies rapidly rotating only inside a restricted central region, significantly smaller than $\sim1\re$. This inner kinematic core is decoupled from the rest of the galaxy, which generally rotates little, similar to classes (a) or (b).
    
    \item[(d) Counterrotating disks:] Galaxies showing rotation extending well beyond $1\re$, with a symmetry axis coinciding with the photometric one but with a reversal of the sense of rotation inside $1\re$. These galaxies can be explained by the presence of two counterrotating disks. The velocity reversal produces a region with nearly zero velocity on the map, where the two opposite velocities counterbalance. When the kinematics is extracted by modeling the line-of-sight velocity distribution (LOSVD) with a single Gaussian, the two opposite velocities appear as an increase of $\sigma$, which appear as two peaks in the velocity dispersion $\sigma$ on opposite sides of the galaxy nucleus.
    
    \item[(e) Disk-like rotation:] Galaxies with regular velocity extending well outside $1\re$ and with the same symmetry axes as the photometry. The velocity field is generally accurately described by that of a thin rotating disk, with a characteristic $\cos\theta$ (with $\theta$ the eccentric anomaly) variation along similar concentric ellipses.
\end{description}
  
\subsection{Intrinsic Shapes}\label{sec:shape}

\begin{figure*}    
    \includegraphics[width=\textwidth]{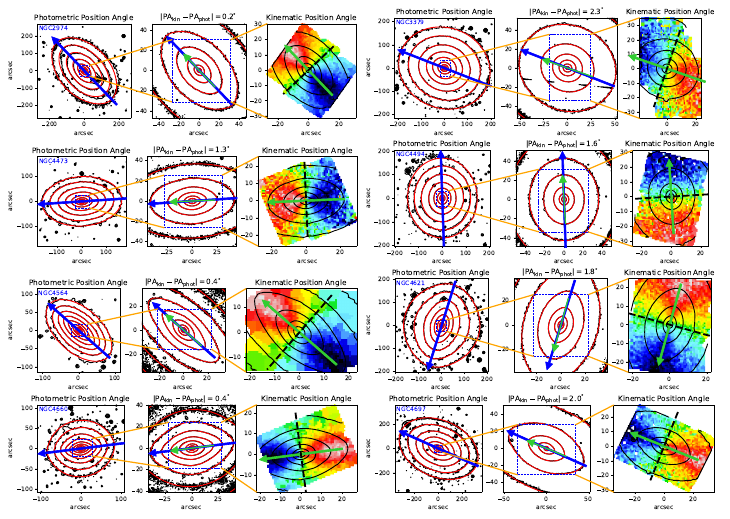} 
    \caption{\textbf{Fast-rotator early-type galaxies (ETGs) are  axisymmetric out to large radii:} For each of the eight galaxies, the left and middle panels show the SDSS surface brightness contours (in black) at two different scales for regular rotator galaxies classified as ellipticals. The red contours represent a Multi-Gaussian Expansion (MGE) fit using \href{https://pypi.org/project/mgefit/}{\texttt{mge.fit\_sectors}} \citep{Cappellari2002mge}, assuming a fixed photometric position angle (PA\textsubscript{phot}) at all radii. The PA\textsubscript{phot}, determined using \href{https://pypi.org/project/mgefit/}{\texttt{mge.find\_galaxy}}, is indicated by blue arrows. The close match between the observed images and the MGE fit confirms that the photometry is consistent with a constant PA\textsubscript{phot} across all radii, with no detectable photometric twists. This behavior contrasts with triaxial systems, where radial ellipticity variations typically cause PA\textsubscript{phot} twists. The right panel displays stellar kinematic maps (from \citealt{Cappellari2011a}) and the best-fitting kinematic position angle (PA\textsubscript{kin}), measured using \href{https://pypi.org/project/pafit/}{\texttt{pafit.fit\_kinematic\_pa}} \citep{Krajnovic2006}. In all cases, the kinematic misalignment is $|\Psi_\text{mis}| \lesssim 2^\circ$, within measurement uncertainties. The absence of PA\textsubscript{phot} variation and the negligible $\Psi_\text{mis}$ indicate that these galaxies maintain axisymmetry out to at least $4\re$, the extent of the photometric data. These eight galaxies are representative of regular-rotator ETGs as a class, unless barred or otherwise disturbed.
        \label{fig:kin_mis_ngc4621}}
\end{figure*}

\begin{figure*} 
    \centering  
    \includegraphics[width=\textwidth]{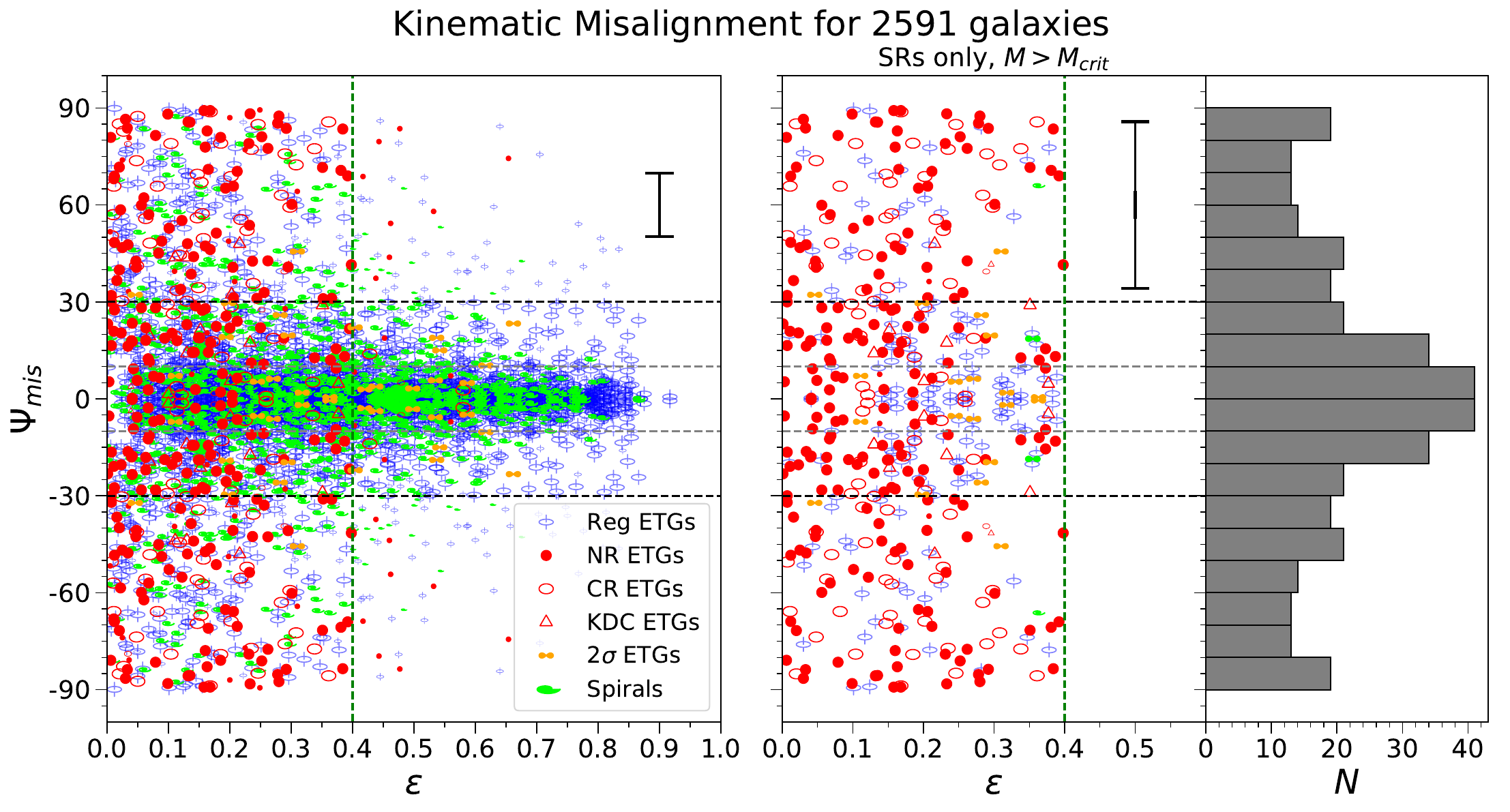}    
    \caption{Left panel: Misalignment ($\Psi_{\rm mis}$) between the photometric axis ($\rm PA_{phot}$) and kinematic axis ($\rm PA_{kin}$) versus ellipticity ($\epsilon$) for galaxies with classifiable kinematics (symbols as in \autoref{fig:kinematic_morphology}). Smaller symbols denote galaxies excluded from the clean sample. Each galaxy is symmetrically duplicated above and below the zero. The vertical dashed green line indicates the $\epsilon < 0.4$ criterion for slow rotators from \citet[eq.~19]{Cappellari2016}. Black horizontal dashed lines mark the $|\Psi_{\rm mis}| = 30^\circ$ and $|\Psi_{\rm mis}| = 10^\circ$ thresholds. The typical error in PA is shown. Right panel: The same selection as the left panel, but only slow rotators with masses exceeding $\mcrit=2\times10^{11}\msun$ are displayed. A histogram of the distribution is included to the right of the scatter plot \citep[fig.~12]{Graham2018}.
    \label{fig:kin_mis}}
\end{figure*}

Measuring the intrinsic shapes of ETGs is challenging because we only observe their projection on the plane of the sky, and they lack the dusty spiral arms that could trace the equatorial plane. Even if we assume ETGs have density distributions approximately stratified on similar triaxial spheroids, with intrinsic axial ratios $p$ and $q$, the two-dimensional distribution $f(p,q)$ of the intrinsic axial ratios cannot be inferred from the observed one-dimensional distribution of observed axial ratios $f(q')$. With IFS observations, we can additionally measure the distribution of the kinematic misalignment angle:
\begin{equation}
    \Psi_{\rm mis} = \mathrm{PA_{kin} - PA_{phot}},
\end{equation} 
which is the difference between the photometric major axis PA\textsubscript{phot} and the kinematic major axis PA\textsubscript{kin}. The definition of these angles is illustrated in \autoref{fig:kin_mis_ngc4621} for eight regular-rotator ETGs (kinematic-morphology class [e] in \autoref{fig:kinematic_morphology}). I measured the PA\textsubscript{phot} with the \href{https://pypi.org/project/mgefit/}{\texttt{find\_galaxy}} function of the \texttt{mgefit} package\footnote{\url{https://pypi.org/project/mgefit/}} \citep{Cappellari2002mge} and the PA\textsubscript{kin} with with the \href{https://pypi.org/project/pafit/}{\texttt{fit\_kinematic\_pa}} function of the \texttt{pafit} package\footnote{\url{https://pypi.org/project/pafit/}} \citep[appendix~C]{Krajnovic2006}.
Note that stellar kinematics can generally be measured to much smaller radii than photometry. However, by comparing PA\textsubscript{kin} within $1\re$ against PA\textsubscript{phot} measured at large radii, we can study the shapes of the outer stellar halos.

Unfortunately, adding the kinematic misalignment angle does not reduce the degeneracy of the problem because we do not know the misalignment angle between the angular momentum vector and the rotation axis of the galaxy. The angular momentum vector can lie anywhere in the plane containing the major and minor axes of the triaxial ellipsoid.

The issue may seem unsolvable in principle. However, when the first statistically significant sample of galaxies was observed, it was found that the kinematic misalignment for regular rotators tended to be small and nearly consistent with zero within measurement errors (see \autoref{fig:kin_mis_ngc4621} and \autoref{fig:kin_mis}). The lack of kinematic misalignment for a significant sample of galaxies, out to their stellar halos, can only occur if these galaxies, as a class, are approximately axisymmetric \citep{Krajnovic2011, Graham2018}. They are as axisymmetric as spiral galaxies \citep{Barrera-Ballesteros2015}. Most observed deviations from axisymmetry are consistent with the presence of bars, disturbances due to mergers, or lower-quality data. A statistical inversion of the intrinsic shapes of fast rotators was performed by \citet{Weijmans2014}.

The situation is dramatically different for non-regular rotators, which tend to be relatively round in projection $\varepsilon\lesssim0.4$. They show almost no preferred kinematic misalignment angle. This is partly due to their near lack of rotation, making PA\textsubscript{kin} extremely unreliable. However, genuine and significant kinematic misalignment is observed in several non-regular rotators, indicating they must be triaxial. Their rather small maximum observed ellipticity indicates that they are at most weakly triaxial and quite close to spherical, with axial ratios $p>q\gtrsim0.6$. A detailed statistical inversion of the shape of slow rotators was done by \citet{Li2018shapes}.

Counter-rotating disks (kinematic-morphology class [d] in \autoref{fig:kinematic_morphology}) are an exception. They are closely related to regular-rotators due to the presence of stellar disks, which explains their flattened shapes.

\subsection{Dynamical Modeling and Orbital Anisotropy}\label{sec:dynamics}

\begin{figure*} 
    \centering
    \includegraphics[width=\textwidth]{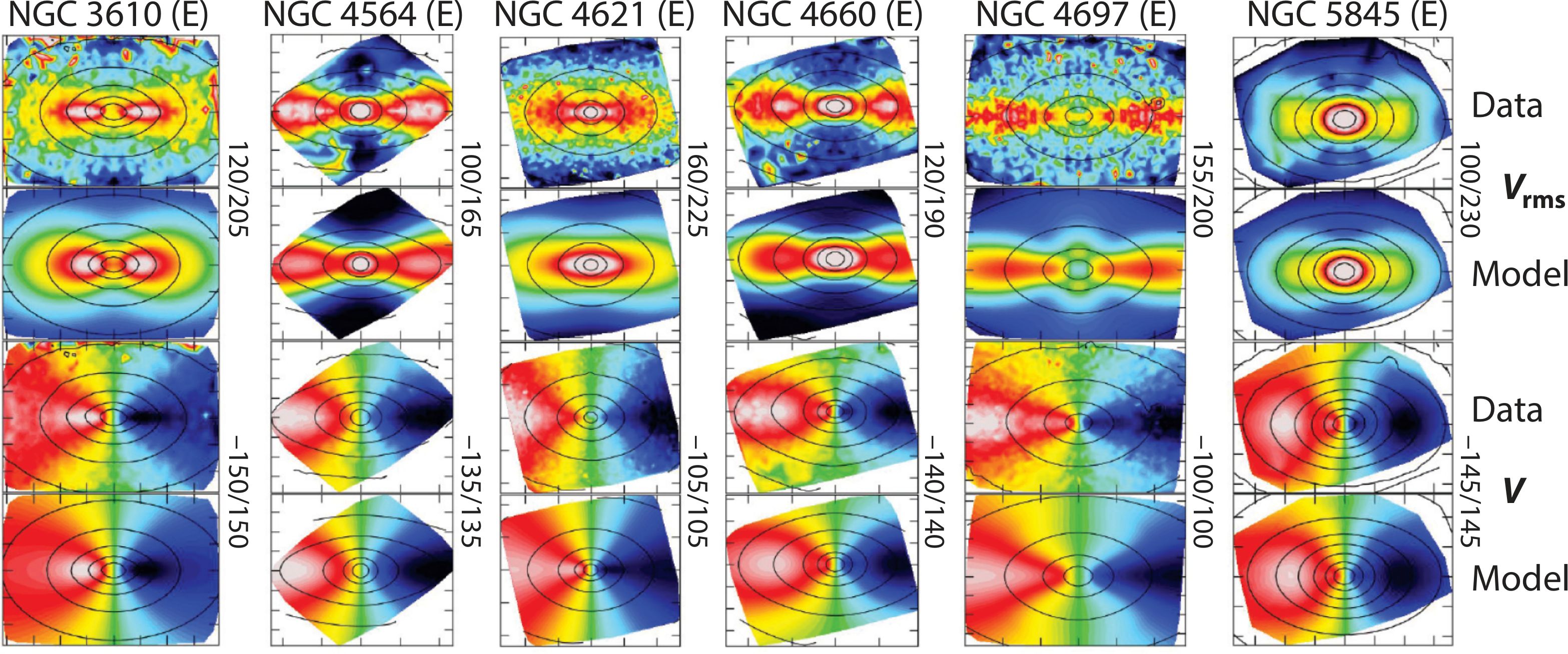}    
    \caption{Jeans Anisotropic Models with a cylindrically aligned velocity ellipsoid (JAM\textsubscript{cyl}) applied to regular rotators morphologically classified as elliptical galaxies. Each plot is divided into four panels: Top: Symmetrized observed stellar $V_{\text{rms}} \equiv \sqrt{V^2 + \sigma^2}$. Second: Best-fitting JAM\textsubscript{cyl} model for $V_{\text{rms}}$. Third: Symmetrized observed mean stellar velocity $V$. Bottom: Best fit to $V$ using the optimal anisotropy $\beta_z$, inclination $i$, and mass-to-light ratio ($M/L$) derived from the $V_{\text{rms}}$ fit, with only an overall scaling applied to $V$. These models accurately predict the shapes of both the $V$ and $V_\text{rms}$ stellar kinematics by adjusting only the anisotropy parameter $\beta_z$ and choosing an inclination, underscoring the dynamical uniformity of regular rotators. Given that the total density in these models is proportional to the stellar density, the results also suggest that the total density slope closely approximates the stellar density within the fitted regions \citep[Fig.~10]{Cappellari2016}. \label{fig:jam_models}}
\end{figure*}

\begin{figure*} 
    \centering
    \includegraphics[width=.49\textwidth]{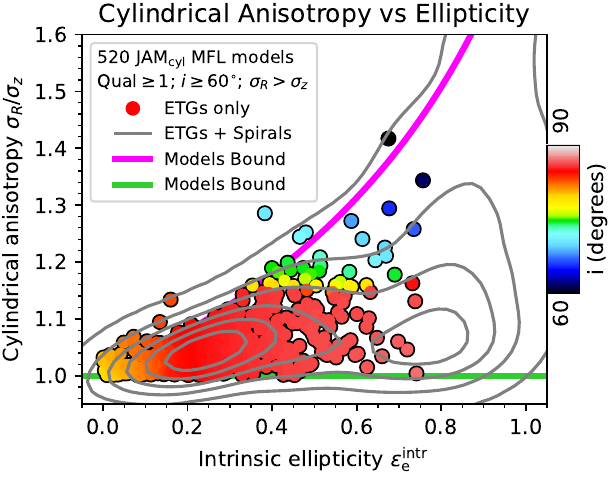}    
    \includegraphics[width=.49\textwidth]{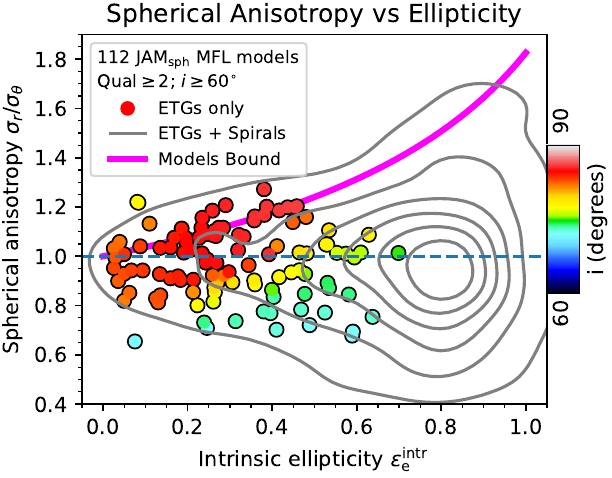}    
    \caption{
        Left panel: The mean anisotropy ratio $\sigma_R/\sigma_z$ within the half-light radius ($R \lesssim \re$) is shown for early-type galaxies (ETGs) from the MaNGA survey. These ETGs, classified morphologically as $T_\mathrm{type} \leq -0.5$ by \citet{VazquezMata2022}, were analyzed using cylindrically-aligned Jeans Anisotropic Models (JAM\textsubscript{cyl}) from \citet{Cappellari2008}. The sample includes galaxies with data quality $\text{Qual} \geq 1$ and inclinations $i > 60^\circ$ to minimize mass-deprojection and inclination-anisotropy degeneracies. The anisotropy ratio is plotted against the intrinsic ellipticity $\eps^\mathrm{intr} \equiv 1 - \sqrt{1 + \eps (\eps - 2)/\sin^2 i}$, where $\eps$ is the observed ellipticity and $i$ is the best-fitting model inclination. Similar results are obtained for $\text{Qual} \geq 2$, but the sample size is too small as I also exclude models on the lower boundary $\sigma_R = \sigma_z$ (green line).        
        Right panel: The mean anisotropy ratio $\sigma_r/\sigma_\theta$ is shown for spherically-aligned models (JAM\textsubscript{sph}) from \citet{Cappellari2020}, selected with $\text{Qual} \geq 2$. The models are constrained to approximately lie below the magenta line (and above the green line for the left panel). For this reason, the distribution of models between these lines should not be interpreted as an empirical trend. However, most models tend to avoid the magenta boundary, with a clear peak near the isotropy line (ratio = 1), especially when including spiral galaxies. Models closest to the magenta boundary, indicating high anisotropy, tend to have lower inclinations, suggesting their anisotropy may be overestimated due to mass-deprojection degeneracies.        
        Gray contours represent a kernel density estimate of the number density for the general population, including both ETGs and spiral galaxies (using \href{https://docs.scipy.org/doc/scipy/reference/generated/scipy.stats.gaussian_kde.html}{scipy.stats.gaussian\_kde}; \citealt{Scipy2020}). Colors indicate the loess-smoothed mean inclination for ETGs only (using \href{https://pypi.org/project/loess/}{loess.loess\_2d}; \citealt{Cappellari2013p20}). The anisotropy distribution of ETGs appears to smoothly extend that of spiral galaxies.        
        These results are based on models where mass follows light (MFL) to reduce mass-anisotropy degeneracy. The total density is close to the true value within the regions ($R \lesssim\re$) sampled by the kinematics. Similar, albeit noisier, results are obtained for models including an NFW halo.        
        Data for this figure were taken from the DynPop catalog \citep[using the keywords \texttt{eps\_mge}, \texttt{inc\_deg}, \texttt{beta\_z}, and \texttt{beta\_r}]{Zhu2023}. Note: The model bounds were determined using the intrinsic ellipticity derived from deprojected MGEs. This measure is not included in the catalog and, for galaxies with variable ellipticity, differs slightly from the $\eps^\mathrm{intr}$ values displayed in the plots.
        }
        \label{fig:anisotropy}
\end{figure*}

\begin{figure*} 
    \centering  
    \includegraphics[width=.6\textwidth]{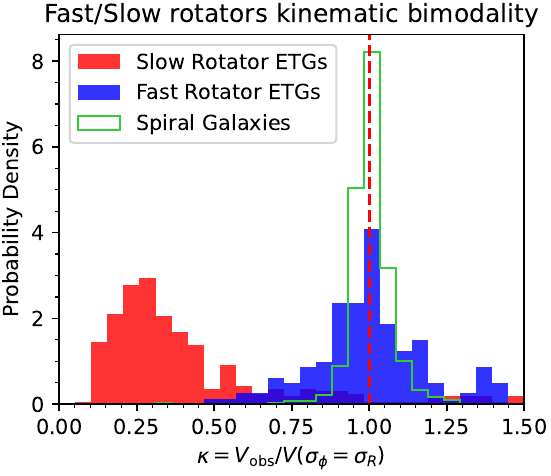}    
    \caption{Histogram of the ratio $\kappa$ between the observed velocity $V_{\rm obs}$ and the velocity $V(\sigma_\phi=\sigma_R)$ predicted by a JAM\textsubscript{cyl} mass-follows-light model with a cylindrically-aligned and oblate ($\sigma_R=\sigma_\phi \neq \sigma_z$) velocity ellipsoid. Similar results are obtained using JAM\textsubscript{sph} models with a spherically-aligned velocity ellipsoid and $\sigma_r=\sigma_\phi \neq \sigma_\theta$. Specifically, $\kappa$ is calculated as the ratio between the projected angular momenta of the data and the model \citep[eq.~52]{Cappellari2008}. The distribution peaks sharply around $\kappa=1$, with an rms scatter that depends on data quality \citep[see][fig.~10]{Zhu2023}. Galaxies were selected with an inclination $i > 60^\circ$ to minimize mass-deprojection and inclination-anisotropy degeneracies. For fast rotators (classified via \autoref{eq:fast_slow_divide}) and spiral galaxies \citep[classified as $T_\mathrm{type} > -0.5$ by][]{VazquezMata2022}, a quality limit of $\text{Qual} \ge 3$ was applied, resulting in $N_\mathrm{fast}=160$ and $N_\mathrm{spiral}=961$ galaxies. For slow rotators, a lower quality threshold of $\text{Qual} \ge 1$ was adopted, yielding $N_\mathrm{gal}=489$ to improve statistics given their smaller fraction. All histograms represent probability densities, i.e., they are normalized to unit area.
    The fact that slow rotators exhibit less rotation than fast rotators is expected by design. However, the clear bimodality, the flat minimum between the two classes, and the clustering of fast rotators around $\kappa=1$, similar to spiral galaxies, are not. This suggests that the two classes of ETGs are physically distinct and do not form a continuous distribution in angular momentum.    
    The data for this figure were obtained from the DynPop catalog by \citet[keyword \texttt{kappa}]{Zhu2023}. \label{fig:kappa}}
\end{figure*}
 
\begin{figure*} 
    \centering
    \includegraphics[width=.49\textwidth]{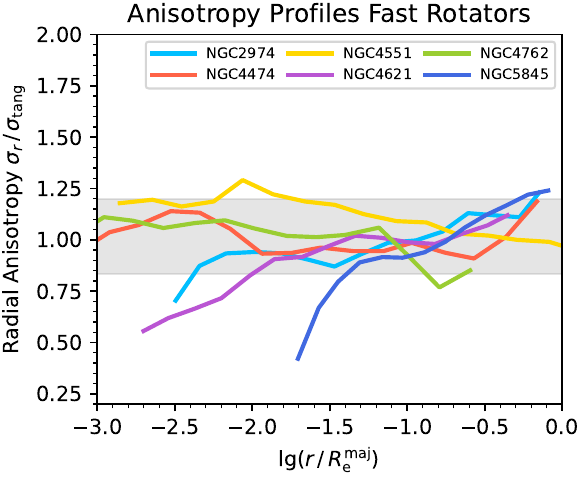}    
    \includegraphics[width=.49\textwidth]{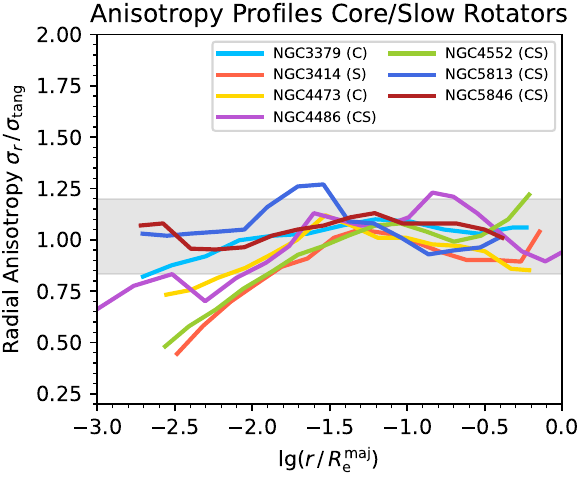}    
    \caption{Radial anisotropy profiles $\sigma_r/\sigma_\mathrm{tang}$ obtained from a subset of Schwarzschild's dynamical models, utilizing high-quality observational data. These profiles are derived from models that have been fitted to both high-resolution nuclear and large-scale ($\sim1\re$) high signal-to-noise ($S/N$) integral field stellar kinematics. In spherical coordinates, $\sigma_r$ denotes the second moment of the velocity distribution along the radial direction, while $\sigma_\mathrm{tang}^2 \equiv (\sigma_\theta^2 + \sigma_\phi^2)/2$ represents the average dispersion in the tangential directions. The anisotropy is defined in the standard way such that $\sigma_\mathrm{tang}$ excludes mean stellar rotation, and an isotropic system with $\sigma_r = \sigma_\theta = \sigma_\phi$ yields $\sigma_r/\sigma_\mathrm{tang} = 1$, irrespective of rotation. Different lines correspond to measurements for distinct galaxies. A gray horizontal band indicates the reference range $5/6 < \sigma_r/\sigma_\mathrm{tang} < 6/5$. The left panel displays models for fast rotators, while the right panel includes slow rotators (as defined by \autoref{eq:fast_slow_divide}, using $\lameps$ from \citealt{Emsellem2011}) or galaxies with a nuclear core in their surface brightness profile (as per \citealt{Krajnovic2013p23}). Parentheses next to galaxy names indicate core galaxies (C), slow rotators (S), or both (CS). 
    Overall, the profiles in both panels are isotropic ($\sigma_r/\sigma_\mathrm{tang} \approx 1$) within $\sim20\%$ for $\re/30 \lesssim r \lesssim \re$. However, slow rotators and/or core galaxies exhibit a systematic trend toward tangential anisotropy within the core radius and near the central supermassive black hole ($r \lesssim \re/100$). Fast rotators do not show a clear systematic trend in their nuclear anisotropy profiles. The uncertainties in the anisotropy profiles are dominated by systematic effects. The profiles are likely intrinsically smooth, implying that the observed $\sim10\%$ fluctuations in individual profiles likely reflect statistical uncertainties.     
    The anisotropy profile for NGC~3379 is sourced from \citet[fig.~15]{Shapiro2006}, while those for NGC~2974, NGC~3414, NGC~4473, NGC~4552, NGC~4621, NGC~5813, NGC~5845, and NGC~5846 are from \citet[fig.~2]{Cappellari2008iaus}. The profile for NGC~4486 is from \citet[fig.~7]{Gebhardt2011}, and those for NGC~4474, NGC~4551, and NGC~4762 are from \citet[fig.~11]{Krajnovic2018}. \label{fig:anisotropy_radial}}
\end{figure*}

For external galaxies (excluding the Milky Way or some very nearby local-group galaxies), we can only measure line-of-sight velocities of their stars. This means the largest amount of dynamical information we can extract from a galaxy is a three-dimensional distribution, specifically the LOSVD at every spatial location on the galaxy image. For a fixed gravitational potential, the dynamics of a steady-state galaxy are fully described by the three-dimensional distribution function (DF), which is a function of the three isolating integrals of motion \citep[e.g.,][sec.~4.2]{Binney2008}. This implies that if (i) we knew the gravitational potential, and (ii) the galaxy was edge-on to avoid the mass deprojection degeneracy, one could theoretically invert the DF from the observed kinematics. However, these conditions are never fully satisfied for real observations of distant galaxies, leading to inherent degeneracies in modeling results.

Three main techniques have been used to model the stellar dynamics of ETGs, to understand their orbital distribution and the distribution of their mass content:

\begin{description}
    \item [Jeans Modeling:] Based on the moments of the collisionless Boltzmann equations, first applied to galaxies by \citet{Jeans1922}. It describes stars like an incompressible fluid in phase space moving under the influence of gravity and computes a prediction for the velocity moments, which are then fitted to the observed kinematics.
    \item [Orbital-Superposition Technique:] Proposed by \citet{Schwarzschild1979}. It computes all possible orbits in a fixed gravitational potential and then finds the linear combination of those orbits that best fits the galaxy kinematics and photometry.
    \item [Made-to-Measure Technique:] Introduced by \citet{Syer1996}. In its most general form, this technique tries to nudge a live non-steady-state N-body model towards fitting the observed kinematics. But no proof of convergence exists in this case. With a fixed gravitational potential it is conceptually similar to Schwarzschild's approach, but less efficient.
\end{description}

The strength of Jeans's method lies in its predictive power. One of its generalizations, the Jeans Anisotropic Modelling \citep[JAM,][]{Cappellari2008,Cappellari2020} method has been applied to large samples of galaxies. It was found that, given the surface brightness distribution and assuming axisymmetry, the large variations in the kinematics of regular-rotator ETGs could be captured to good accuracy by just a couple of parameters: the inclination $i$ and the global anisotropy, as illustrated in \autoref{fig:jam_models}. JAM was found to be more accurate in estimating galaxy densities than Schwarzschild's technique for regular rotators, using both real data \citep[fig.~8]{Leung2018} and N-body simulations \citep[fig.~4]{Jin2019}.

Anisotropy can be defined as the ratio between the velocity dispersion measured along different directions by a hypothetical observer in the galaxy moving with the average stellar velocity. It can be measured in different coordinate systems, such as cylindrical polar $(R,\phi,z)$ or spherical polar $(r,\theta,\phi)$. In axisymmetry, the second velocity moments of the line-of-sight velocity $\overline{V^2_\mathrm{los}}$ are independent of the anisotropy in the tangential ($\phi$) direction. This is because one can always reverse the sense of rotation of galaxy stellar orbits without changing the gravitational potential or the orbital makeup while keeping $\overline{V^2_\mathrm{los}}$ unchanged. Therefore, the key anisotropy parameters are: (i) in cylindrical coordinates, the ratio of the velocity dispersion in the axial and radial directions $\sigma_z/\sigma_R$, and (ii) in spherical coordinates, the ratio of the velocity dispersion in the angular and radial directions $\sigma_\theta/\sigma_r$.

The global anisotropy distribution, extracted from the largest set of JAM dynamical models constructed so far \citep{Zhu2023}, is shown in \autoref{fig:anisotropy}. For this figure, I selected a subset of galaxies with the best data and with fitted inclination $i>60^\circ$ to reduce the effect of the light deprojection degeneracy. It shows that the anisotropy of ETGs strongly depends on ellipticity, with larger variations in anisotropy for intrinsically flatter galaxies, as first pointed out by \citet{Cappellari2007, Thomas2009}. However, in general, deviations from isotropy (i.e., equal dispersion in all coordinates) are rather small, typically $\lesssim20\%$. \citet{Santucci2022} presented consistent results using Schwarzschild's technique. However, the extra generality of the method, compared to JAM, results in much larger uncertainties and a correspondingly larger scatter.

\autoref{fig:kappa} shows, for the same galaxy sample, the ratio $\kappa=V_{\rm obs}/V(\sigma_\phi=\sigma_R)$ between the observed mean stellar velocity and the velocity of a JAM model with an oblate velocity ellipsoid ($\sigma_\phi=\sigma_R$). For regular rotator ETGs, this ratio is close to unity with high accuracy \citep[see][fig.~11]{Cappellari2016}. This implies that, once the gravitational potential has been determined, by fitting the $V_\mathrm{rms}$, the rotation of fast rotators can be predicted with high accuracy by assuming $\sigma_\phi=\sigma_R$. This further highlights the remarkable regularity of the kinematics of this class of galaxies.

Schwarzschild's orbit-superposition method leverages the full information content of the line-of-sight velocity distribution (LOSVD) when high-quality kinematic data are available, enabling non-parametric recovery of the orbital structure. This approach has also been adapted to study stellar population properties \citep[e.g.,][]{Poci2019}. Radial anisotropy profiles derived for early-type galaxies (ETGs) with integral field spectroscopy (IFS) are shown in \autoref{fig:anisotropy_radial}, where $\sigma_r$ denotes the radial velocity dispersion in spherical coordinates and $\sigma_\mathrm{tang}^2 \equiv (\sigma_\theta^2 + \sigma_\phi^2)/2$ represents the azimuthally averaged tangential dispersion. These profiles typically transition from tangential anisotropy ($\sigma_\mathrm{tang} > \sigma_r$) in the central regions to radial anisotropy ($\sigma_r > \sigma_\mathrm{tang}$) at larger radii \citep{Gebhardt2003, Cappellari2008iaus, Thomas2014}. Around $\re$ deviations from isotropy are generally mild ($\lesssim20\%$). However, constraints on anisotropy at $r\gtrsim\re$ remain unreliable due to degeneracies with dark halo modeling and sensitivity to systematic biases in low signal-to-noise data.  

\citet[fig.~2]{Thomas2014} performed a similar comparison between anisotropy profiles for core and core-less ETGs. Their result is qualitatively similar, with more homogeneous profiles for core galaxies than core-less ones. However, they find stronger anisotropy gradients for core galaxies. They attribute this to their inclusion of a dark-matter halo in their models. Dark matter makes the total density profile shallower, which in turn makes the anisotropy profile steeper, for fixed kinematics. Of the models in \autoref{fig:anisotropy_radial}, only the model for NGC~4486 includes a dark-matter halo. The other models assume the total density follows the light distribution, which is a good approximation for the regions sampled by kinematics (\autoref{fig:total_slope}). Although the inclusion of a dark halo would generally be preferable, there is degeneracy between anisotropy and the total density slope \citep{Binney1982,Gerhard1993}. This could be lifted with high $S/N$ integral-field kinematics allowing one to measure the full LOSVD. However, the large-scale data used by \citet{Thomas2014} are mostly based on sparse coverage and modest quality \citep{Rusli2013}. It is unclear whether they can reliably break the mass-anisotropy degeneracy. This is why I did not include these profiles in \autoref{fig:anisotropy_radial}. Models including a dark halo, but with data quality like the models of \autoref{fig:anisotropy_radial} would be extremely valuable.

\subsection{Total Density Profiles and Dark Matter Distribution}
\label{sec:total_density}

\begin{figure*} 
    \centering
    \includegraphics[width=.49\textwidth]{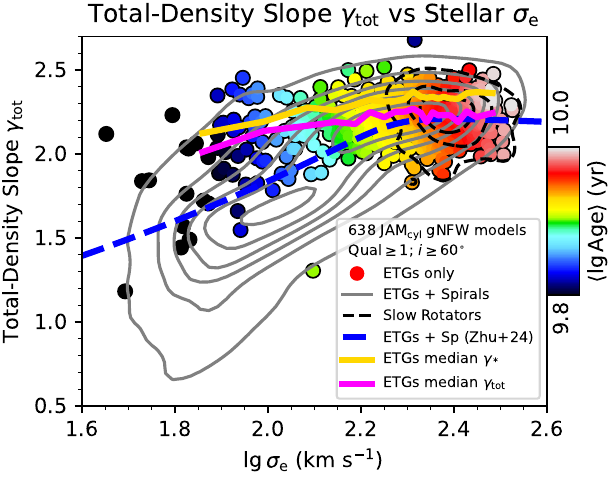}    
    \includegraphics[width=.49\textwidth]{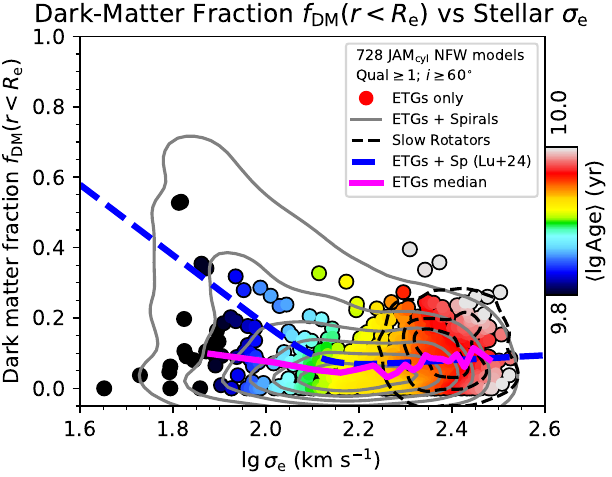}    
    \caption{
    Left panel: The mass-weighted total logarithmic density slopes $\gamma_\mathrm{tot}$ of \autoref{eq:slope} within a sphere of radius $r=\re$, as a function of galaxy effective velocity dispersion $\se$, for the most general JAM\textsubscript{cyl} models with gNFW halo. The filled circles represent ETGs. They exhibit steeper total slopes at fixed $\se$ compared to the overall galaxy population because they are generally older and follow the age-slope trend at fixed $\se$ \citep[figs.~8 and 10]{Zhu2024}. The blue dashed line is the best-fitting relation to the median of the whole population from \citet[eq.~13]{Zhu2024}. The gold solid line shows the median stellar luminosity density slope $\gamma_*$ for the ETGs, computed in 20 bins of $\se$ with an equal number of galaxies, and tracks $\gamma_\mathrm{tot}$ with a constant offset \citep[fig.~1]{Li2024DynPop6}.        
    Right panel: The correlation between the dark matter fraction $f_{\rm DM}(\re)$ within a sphere of radius $1\re$  and $\se$, from JAM\textsubscript{cyl} models with an NFW halo. The blue dashed line represents the fit to the median values for the overall population, which remains nearly flat for $\lg(\se/\kms)\gtrsim2.1$ and increases at lower $\se$ \citep[eq.~7]{Lu2024}. ETGs follow the overall trend at high $\se$ but exhibit lower $f_\mathrm{DM}(\re)$ at lower $\se$, consistent with their steeper total density profiles in the left panel, primarily due to the steeper stellar density (larger bulge mass fraction) than spirals. The low $f_{\rm DM}(\re)$ for ETGs can also be inferred non-parametrically from \autoref{eq:fdm_from_slopes}, using the values of $\gamma_\mathrm{tot}$ and $\gamma_*$ in the left panel, assuming $\gamma_\mathrm{DM}=1$.        
    \emph{In both panels}, the filled circles represent ETGs \citep[classified as $T_\mathrm{type}\le-0.5$ by][]{VazquezMata2022}, colored by the loess-smoothed global luminosity-weighted age (using \href{https://pypi.org/project/loess/}{loess.loess\_2d}; \citealt{Cappellari2013p20}). The magenta solid line indicates the median value for the ETGs alone, computed in 20 bins of $\se$ with an equal number of galaxies. The gray contours show the kernel density estimate of the density distribution for the entire galaxy population of ETGs and spiral galaxies (using \href{https://docs.scipy.org/doc/scipy/reference/generated/scipy.stats.gaussian_kde.html}{scipy.stats.gaussian\_kde}; \citealt{Scipy2020}), while the black dashed contours is the same density for slow rotators alone. The data for this figure come from the DynPop catalog: the dynamical quantities from \citet[keywords \texttt{MW\_Gt\_Re}, \texttt{MW\_Gs\_Re}, \texttt{Sigma\_Re}]{Zhu2023}, and the stellar ages from \citet[keyword \texttt{LW\_Age\_Re}]{Lu2023}.
    \label{fig:total_slope}}
\end{figure*}

The primary quantity that dynamical models can extract from kinematic data is the total density profile. Its robustness comes from the fact that the total density does not depend on assumptions about how to parametrize the luminous and dark matter components separately. In particular, this quantity is independent of possible gradients in the stellar population in galaxies.

Modeling the gravitational lensing of 58 ETGs revealed that, within the region $R\lesssim\re/2$ constrained by the lenses, the density follows $\rho(r)\propto r^{-\gamma_\mathrm{tot}}$ with (positive) logarithmic density slope $\gamma_\mathrm{tot}\equiv-d\lg\rho(r)/d \lg r$ given by $\gamma_\mathrm{tot}=2.08\pm0.02$ \citep{Koopmans2009,Auger2010}. This is close to the isothermal slope $\gamma_\mathrm{tot}=2$, which corresponds to flat circular velocity curves. Dynamical models of the stellar kinematics of 14 regular-rotator ETGs, extending out to much larger radii of $\approx4\re$, found a slightly steeper slope $\gamma_\mathrm{tot}=2.19\pm0.03$ over the whole radial range, with equally small scatter \citep{Cappellari2015dm}.

This apparent `universal' slope of the density profiles later turned out to be valid only for galaxies with large effective velocity dispersion $\se$. The slope is nearly constant, with little scatter, at $\gamma_\mathrm{tot}\approx2.2$ for $\lg(\se/\kms)\gtrsim2.2$ but starts decreasing at lower $\se$ (see fig.~22 of \citealt{Cappellari2016}, and \citealt{Poci2017}). The trend is weak in ETGs, but becomes more apparent when including both spiral galaxies and ETGs, as the mean slopes for spirals reach $\gamma_\mathrm{tot}\approx1.5$ at $\lg(\se/\kms)\approx1.7$ \citep{Li2019, Zhu2024,Li2024DynPop6}. The left panel of \autoref{fig:total_slope} shows the $\gamma_\mathrm{tot}-\se$ relation for ETGs as a continuation of a smooth trend linking spirals and ETGs. The total slope $\gamma_\mathrm{tot}$ is lower for younger galaxies and tends to be systematically higher for ETGs, which have the oldest stellar populations \citep{Zhu2024,Li2024DynPop6}.

To study the separate contributions of luminous and dark matter, which are key elements of our current galaxy-evolution paradigm, one must make additional assumptions: (i) adopt a parametrization, and possibly a shape, for the dark matter distribution; and (ii) decide how to extract and approximate the baryonic matter from observations.

The reliance on extra assumptions makes the results on dark matter distribution less robust than those on the total slopes. For example, if the dark matter profile approximates the stellar one and the total density is close to the stellar density, a change in the stellar $M/L$ becomes indistinguishable from a change in the dark matter fraction $f_\mathrm{DM}(\re)$, using purely dynamical or strong gravitational lensing techniques. This situation frequently occurs in real ETGs, as the total slope is typically close to the stellar one (\citealt{Cappellari2013p15,Poci2017,Li2024DynPop6}; \autoref{fig:total_slope}).

The stellar mass distribution is inferred by deprojecting the observed surface brightness. This process is mathematically non-unique, even if the galaxy is assumed axisymmetric with a known inclination, unless the galaxy is edge-on \citep{Rybicki1987,Gerhard1996}. The deprojection becomes much more degenerate for any viewing direction if the galaxy is assumed triaxial. This deprojection degeneracy is a major uncertainty in any dynamical modeling technique. Additionally, gradients in the stellar $M/L$, inferred from the galaxy's stellar population or colors, must be included in the models, increasing the resulting uncertainties of the dark matter decomposition.

A common description for the dark matter distribution is the generalized Navarro, Frenk, and White (gNFW) profile \citep{Wyithe2001}:
\begin{equation}
    \rho_(r)=\rho_s \left(\frac{r_s}{r}\right)^\gamma
        \left(\frac{1}{2}+\frac{1}{2}\frac{r}{r_s}\right)^{\gamma-3}.
\end{equation}
The density has the same large-radii asymptotic power-law slope $\rho(r)\propto r^{-3}$ as the NFW halo \citep{Navarro1996nfw}, but it allows for a variable inner slope $\gamma$. The ranges include a flat inner core ($\gamma=0$) and the NFW slope ($\gamma=1$) as special cases. The gNFW profile is often used to describe the total density rather than just the dark matter.

Early studies on small samples of galaxies indicated that the dark matter fraction $f_{\rm DM}(\re)$ in ETGs contributes minimally to the total mass within $1\re$ \citep[e.g.,][]{Gerhard2001}. A more extensive study involving 260 galaxies using IFS data and JAM models confirmed these findings, showing that the median $f_{\rm DM}(\re)$ was as low as 13\% \citep{Cappellari2013p15}. Recently, \citet{Lu2024} extended this analysis using JAM on a sample of over 2000 ETGs and more spirals. After carefully removing unreliable models, they reported a clear relationship between $f_{\rm DM}(\re)$ and $\se$, as shown in the right panel of \autoref{fig:total_slope}. Their median trend is flat for $\lg(\se/\kms)\gtrsim2.1$ (close to where the $\gamma_\mathrm{tot}-\se$ relation flattens), with $f_{\rm DM}(\re)\approx9\%$. 

Here, I repeated their analysis for the ETGs subsample alone. Following their methodology, I first used the $(M_*/L)-\se$ relation determined from JAM models to exclude likely unreliable models. For this, I employed the robust linear fitting method implemented in the \texttt{ltsfit} package\footnote{\url{https://pypi.org/project/ltsfit/}} described in \citet{Cappellari2013p15}, which combines the Least Trimmed Squares robust technique of \citet{Rousseeuw2006} with a least-squares fitting algorithm that accounts for errors in all variables, intrinsic scatter, and automatic outlier detection.

I found that the relationship for ETGs is an extension of a trend including spirals, with ETGs showing systematically lower $f_{\rm DM}(\re)$ than spirals at a given $\se$. The dark matter density profiles show a preference for being steeper than the NFW slope ($\gamma=1$) \citep{Li2024DynPop6}, as predicted by the adiabatic contraction effect of baryons on dark matter particles during gas accretion \citep[e.g.,][]{Gnedin2004}. 
Broadly consistent results were presented by \citet{Santucci2022} using Schwarzschild's method, though with much larger scatter due to the more flexible models, making it harder to detect any trend.

The trends in dark matter fraction can be understood without the need to assume any specific dark halo parametrization. When the logarithmic density slopes are mass-weighted, as shown in \autoref{fig:total_slope}, and defined by \citep[eq.~1]{Dutton2014}:
\begin{equation}\label{eq:slope}
    \langle\gamma(r)\rangle 
    \equiv \frac{1}{M(r)}\int_{0}^{r}\gamma(x)4\pi x^2\rho(x)dx 
    = \frac{1}{M(r)}\int_{0}^{r}-\frac{d\lg\rho(x)}{d\lg r} 4\pi x^2\rho(x)dx 
    = 3 - \frac{4\pi r^3\rho(r)}{M(r)},
\end{equation}
there is a very simple yet rigorous relationship between the mass-weighted total slope $\gamma_\mathrm{tot}$, dark halo slope $\gamma_\mathrm{DM}$, stellar slope $\gamma_*$, and the dark matter fraction within the same radius. This relationship is independent of the assumed radial density profiles of the stars and dark matter:
\begin{equation} \label{eq:fdm_from_slopes}
    f_\mathrm{DM} = \frac{\gamma_\mathrm{tot}- \gamma_*}{\gamma_\mathrm{DM} - \gamma_*}.
\end{equation}
It can be used to infer the dark matter fraction from the total and stellar slopes, which can be measured without degeneracies, while only assuming a mass-weighted dark-matter slope.

Assuming the mean dark-halo slope within $1\re$ follows the NFW profile with $\gamma_\mathrm{DM} = 1$, and adopting the nearly constant median offset $\gamma_* - \gamma_\mathrm{tot} = 0.11$ that I measured for the ETG subsample (\autoref{fig:total_slope} left), I derived from \autoref{eq:fdm_from_slopes} that $f_\mathrm{DM}$ ranges from 8\% to 10\% when $\gamma_\mathrm{tot}$ varies between 2.2 and 2.0. This simple estimate agrees with the detailed models shown in the right panel of \autoref{fig:total_slope}. One can also use \autoref{eq:fdm_from_slopes} to estimate the effect of possible halo contraction without running new models: For a maximally contracted dark halo with $\gamma_\mathrm{DM} \approx 1.6$ \citep[fig.~2]{Cappellari2013p15}, the median dark matter fraction within $\re$ in ETGs would vary between 15\% and 22\%.

\subsection{Quantitative Kinematic Classification}\label{sec:lam_eps}

\begin{figure*}   
    \centering
    \includegraphics[height=.4\textwidth]{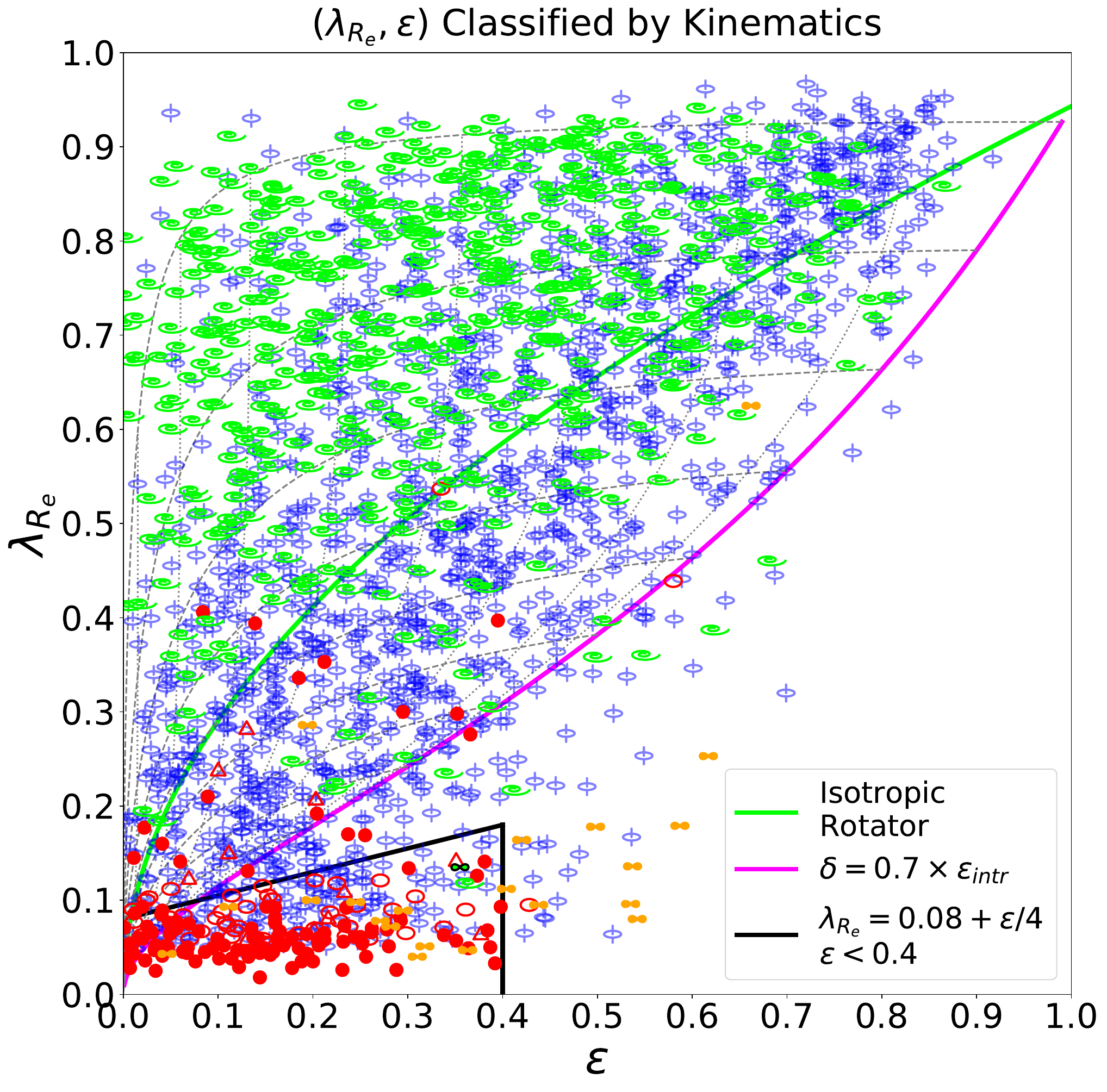}    
    \includegraphics[height=.4\textwidth]{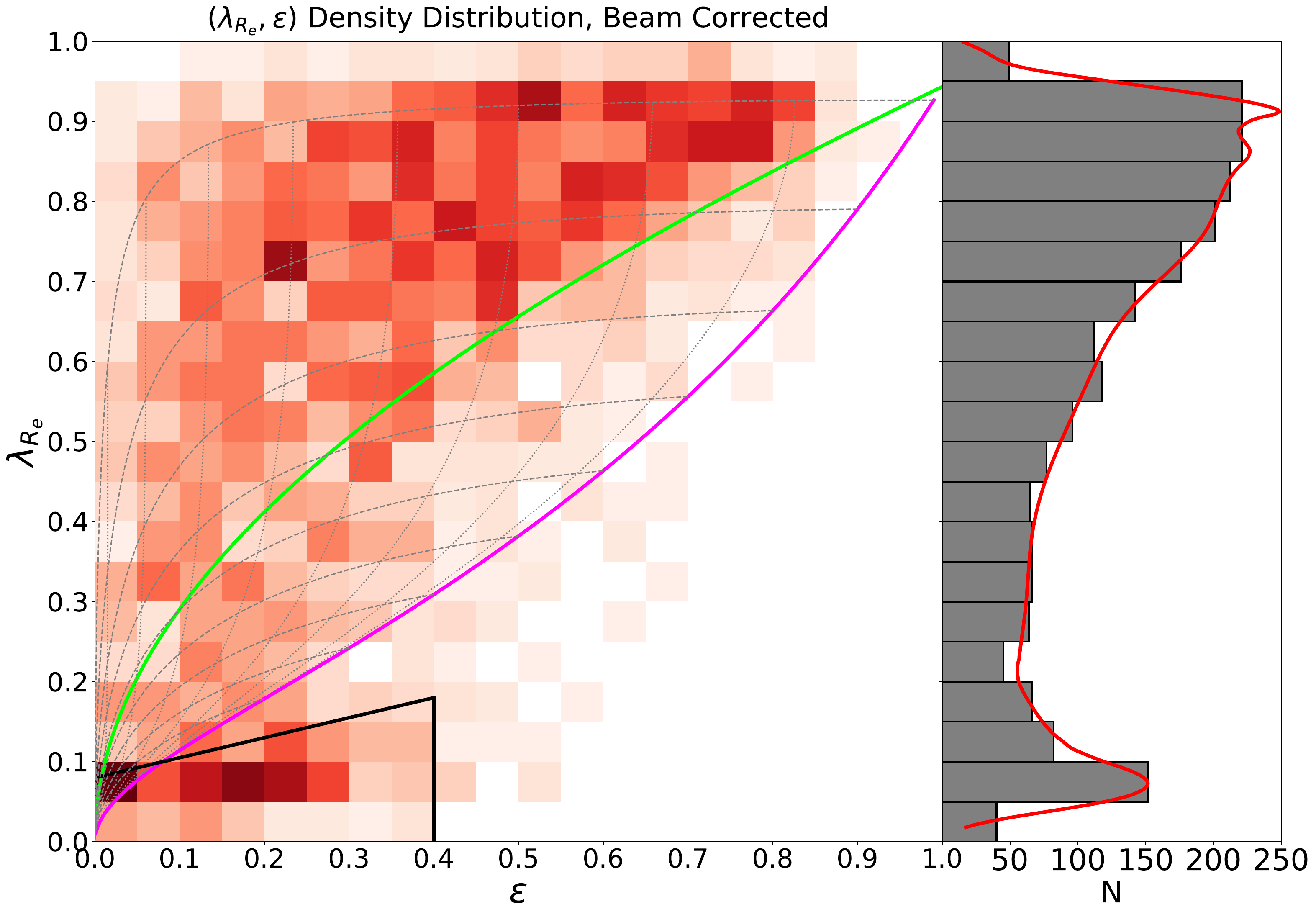}    
    \caption{Left panel: the $\lameps$ diagram is labeled according to kinematic morphology, with symbols representing ETGs as described in \autoref{fig:kinematic_morphology}, and spirals marked in green. The points represent the observed $\lameps$ values for 2,286 galaxies from the MaNGA survey, obtained through integral-field spectroscopy. The thick green line is the prediction for an edge-on ($i = 90^\circ$) isotropic rotator from \citet{Binney2005}, approximately converted from $V/\sigma$ into $\lam$ \citep[eq.~B1]{Emsellem2011}, and the magenta line is the converted edge-on anisotropy-ellipticity relation $\delta=0.7\times\eps^\mathrm{intr}$ from \citet{Cappellari2007}. Thin dotted lines show the magenta line at different inclinations ($\Delta i = 10^\circ$), while black dashed lines trace how galaxies with a particular $\epsilon_{\rm intr}$ at $i = 90^\circ$ move with changing inclination, using the formalism from \citet[sec.~3.5, 3.6]{Cappellari2016}. The black trapezium in the lower-left corner define the region for slow rotator ETGs: $\lam < 0.08 + \eps/4$, $\eps < 0.4$ \citep[eq.~19]{Cappellari2016}. 
    Right panel: Same as in the left panel, but with the distribution of galaxies shown as a 2-dim histogram, and as a 1-dim histogram on the right. There is a clear peak in the galaxy distribution inside the trapezium defining the slow rotators \citep[fig.~5]{Graham2018}.
    \label{fig:lam_eps}}
\end{figure*}

\begin{figure*}   
    \centering
    \includegraphics[height=.5\textwidth]{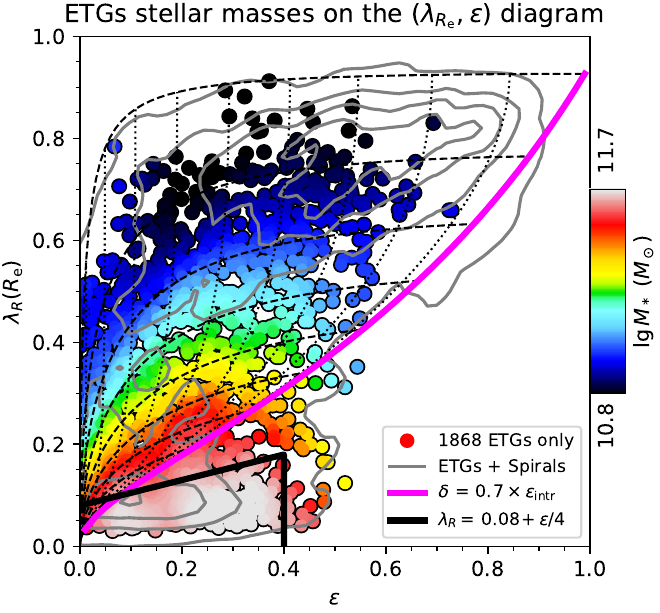}    
    \caption{
        Distribution of dynamically determined ETGs stellar masses on the $\lameps$ diagram. The masses are computed as $M_* \equiv (M/L)_\mathrm{JAM} \times L$ using the JAM\textsubscript{cyl} mass-follow-light models with $\text{Qual}\ge1$, leading to a sample of $N_\mathrm{ETG}=1902$ ETGs. The magenta line represents the edge-on theoretical relation $\delta = 0.7 \times \epsilon_{\rm intr}$ \citep{Cappellari2007}, which approximates the lower boundary of the region of the diagram enclosing most of the fast rotator and spiral galaxies. The dotted lines show how the relation transforms at different inclinations using equations from \citet[sec.~3.5, 3.6]{Cappellari2016}.         
        There are no ETGs flatter than $\eps \gtrsim 0.6$ as those tend to be morphologically misclassified as spirals, though flatter ETGs do exist (e.g., \autoref{fig:atlas3d_comb}). Contours of equal mass follow the dashed lines as predicted, indicating how galaxies of given intrinsic ellipticity $\eps^\mathrm{intr}$ project at different inclinations. This projection is expected if ETGs of increasing mass have a characteristic $\eps^\mathrm{intr}$ (due to a larger bulge) and are projected at random orientations. Massive ETGs tend to lie in the black trapezium of slow rotators \citep[eq.~19]{Cappellari2016}. The data for this figure come from the DynPop catalog by \citet[keywords \texttt{Lambda\_Re}, \texttt{Eps\_MGE}, \texttt{Lum\_tot\_MGE}, \texttt{log\_ML\_dyn}]{Zhu2023}. A version of this figure with both ETGs and spirals was shown in \citet[fig.~15]{Lu2023}.         
        The filled disks represent ETGs, colored by the loess-smoothed global luminosity-weighted age (using \href{https://pypi.org/project/loess/}{loess.loess\_2d}; \citealt{Cappellari2013p20}). The gray contours show the kernel density estimate of the density distribution for the entire galaxy population of ETGs and spiral galaxies (using \href{https://docs.scipy.org/doc/scipy/reference/generated/scipy.stats.gaussian_kde.html}{scipy.stats.gaussian\_kde}; \citealt{Scipy2020}).
        \label{fig:lam_eps_mass}}        
\end{figure*}

The results from the shape distribution in \autoref{sec:shape} and dynamical modeling in \autoref{sec:dynamics} demonstrate that ETGs are not a homogeneous class of objects. Instead, they can be separated into two distinct classes, showing a bimodal distribution in some key parameters:
\begin{description}
    \item [Regular rotators:] These galaxies show small kinematic misalignment and can be modeled as axisymmetric galaxies with stellar disks seen at various inclinations. They have homogeneous orbital makeup, display a range of anisotropies, and have highly predictable stellar rotation velocities or angular momenta.
    \item [Non-regular rotators:] These galaxies are radically different. They are not axisymmetric but weakly triaxial and close to spherical, lacking stellar disks. Their rotation is generally negligible, except for possible kinematically-decoupled cores.
\end{description}

In two companion papers \citep{Emsellem2007,Cappellari2007}, the specific angular momentum parameter $\lambda_R$ was introduced to separate these two kinematic classes using IFS kinematics in an (nearly) inclination-independent manner. This parameter provides a simple and reproducible way to classify galaxies:
\begin{equation}
    \lambda_R\equiv
    \frac{\langle R |V| \rangle}{\langle R \sqrt{V^2+\sigma^2} \rangle}
    =\frac{\sum_{n=1}^N F_n R_n |V_n|}{\sum_{n=1}^N F_n R_n \sqrt{V_n^2+\sigma_n^2}},
    \label{eq:lambda_r}
\end{equation}
Here, $F_n$ represents the fluxes within the $N$ spatial bins where the mean stellar velocities $V_n$ and velocity dispersions $\sigma_n$ are measured. The summation extends to a specific finite radius $R_{\rm max}$ within a galaxy's isophote. A common benchmark is to use $R_{\rm max}=\re$ and denote the parameter as $\lam$.

The $\lambda_R$ parameter is closely related to the $V/\sigma$ rotation parameter \citep{Binney2005}, which is instrumental in analyzing the interplay between shape, rotation, and anisotropy in ETGs. However, a significant distinction lies in the $\lambda_R$ parameter's sensitivity to the spatial distribution of velocities, unlike $V/\sigma$. Initially, the $\lambda_R$ parameter was derived by substituting the velocity with the magnitude of the luminosity-weighted average projected angular momentum, expressed as $\langle\mathbf{L}\rangle = \langle \mathbf{R} \times \mathbf{V} \rangle$. To simplify computation and eliminate the need for determining vector directions, this was further simplified to $\langle R\,|V| \rangle$, producing equivalent results, where $R$ represents the projected distance from the galaxy center. By making this proxy for angular momentum dimensionless and normalizing it with a quantity such as $V^2_{\rm rms} \equiv V^2 + \sigma^2$, which is proportional to the galaxy mass according to the scalar virial theorem \citep[sec.~4.8]{Binney2008}, the $\lambda_R$ parameter is obtained \citep{Emsellem2007}.

Similar to the $V/\sigma$ parameter, the $\lambda_R$ kinematic proxy must be analyzed as a function of the galaxy's apparent ellipticity ($\epsilon$), both measured within the $1\re$ isophote. The $\lameps$ diagram has been systematically applied to growing samples of early-type galaxies (ETGs): \citet{Emsellem2007} pioneered its use for 48 ETGs in the SAURON survey; \citet{Emsellem2011} extended it to 260 galaxies in ATLAS\textsuperscript{3D}; \citet{FalconBarroso2019} to 600 galaxies in CALIFA; \citet{Graham2018} to 2,300 galaxies in MaNGA; and \citet{vandeSande2021} to 3,000 galaxies in SAMI. The largest application to date leverages the final MaNGA survey sample of 10,000 galaxies \citep{Zhu2023}, whose unprecedented size and data quality establish it as the current benchmark for global galaxy property studies. This work predominantly adopts MaNGA-derived figures due to its statistical robustness and kinematic quality.

\autoref{fig:lam_eps} shows the $\lameps$ version by \citet{Graham2018}, which includes the kinematic visual classification of \autoref{fig:kinematic_morphology}. The magenta line represents the edge-on theoretical anisotropy vs. ellipticity relation $\delta=0.7\times\eps^\mathrm{intr}$ \citep{Cappellari2007}, which approximately traces the lower envelope of the distribution of regular rotator ETGs (and spiral galaxies). This region contains disks with varying bulge fractions seen at random orientations. The bulge-free disks and spiral galaxies populate the top of the diagram, with more bulge-dominated systems having lower $\eps$ near the bottom of the magenta-line envelope. This lower envelope's existence can be attributed to an approximate limit to equilibrium solutions for the stellar orbits in galaxies with a nearly-oblate velocity ellipsoid \citep{Wang2021}. The black trapezium defines the new fast and slow rotator classification, quantitatively separating the two classes described at the beginning of this section. It is defined by \citep[eq.~19]{Cappellari2016}:
 \begin{equation}
    \lam < 0.08 + \eps/4 \qquad \text{with} \qquad \eps < 0.4.
    \label{eq:fast_slow_divide}
\end{equation}

Most of the non-regular rotators visually classified with the scheme of \autoref{fig:kinematic_morphology} fall within the slow-rotator region. Counterrotating disks can have $\lam$ values as low as slow rotators but are often significantly flatter. For $\eps < 0.4$, the division in \autoref{eq:fast_slow_divide} is very close to that by \citet{Emsellem2011}, but it introduces a new limit in ellipticity to reduce the contamination of counterrotating disks, which are a physically distinct class of galaxies.

The middle panel of \autoref{fig:lam_eps} shows an approximate limit at $\lam \approx 0.1$, separating fast and slow rotator ETGs naturally emerging from the data bimodality. The MaNGA sample used to construct the $\lameps$ diagram was selected to be flat in stellar mass. However, a histogram of the $\lam$ values reveals a secondary peak within the slow-rotator trapezium region. A sharp transition in stellar masses is also observed between galaxies inside and outside the slow rotator trapezium. \autoref{fig:lam_eps_mass} indicates that galaxies in the slow rotator box generally have larger masses than fast rotators of similar $\eps$. Additionally, flat ($\eps \gtrsim 0.4$) galaxies below the magenta line have much smaller masses than slow rotators at lower ellipticity and are mainly spiral galaxies (see \citealt[fig.~15]{Lu2023}, though not shown in \autoref{fig:lam_eps_mass}). This suggests that flat slowly rotating galaxies are distinct from the massive $M_* \gtrsim \mcrit \approx 2 \times 10^{11} \msun$ slow rotator class. A statistical approach confirming bimodality in the $(M_*, \lam)$ plane is presented by \citet[fig.~7]{vandeSande2021}.

The physical significance of the fast/slow rotator ETG bimodality is supported by the correlation between specific stellar angular momentum and the presence of cores in surface brightness profiles (\autoref{sec:inner_slopes}). Independent studies show overlap between slow rotators and cores in the central surface brightness \citep{Lauer2012, Krajnovic2013p23}.

The low mass of galaxies below the magenta line represents a significant empirical trend. A study of $M_* \approx 10^9 \msun$ galaxies in the MaNGA survey found that these low-mass galaxies, despite kinematic misalignment consistent with axisymmetry like more massive fast rotator ETGs, exhibit systematically lower specific angular momentum \citep{Wang2024}. However, as shown in \autoref{fig:mass-size}(e), their angular momentum is not as low as that of massive slow rotators with $M_* \gtrsim\mcrit$. The low $\lam$ in these galaxies may be due to easier perturbation during their early gas-rich formation phase \citep[e.g.,][]{BlandHawthorn2024} or the tendency of accreting gas disk spins to flip in less than an orbital time due to mergers and changes in cosmic-web streams at low masses \citep{Dekel2020}, leading to significant stellar counter-rotation in the remnant.

Having established the physical distinction between fast and slow rotator ETGs, it is instructive to compare this classification against Hubble's classic morphological classification. One finds that \emph{just about 1/3 of galaxies morphologically classified as ellipticals are slow rotators} \citep{Emsellem2011}. This implies that Hubble's ellipticals classification is generally not physically meaningful, because in the vast majority of cases, it selects fast rotator ETGs. These are inclined galaxies with disks, which would be classified as either S0 or disky ellipticals if seen edge-on. This highlights the usefulness of using the new kinematic classification of ETGs.

\section{Scaling Relations}

\subsection{Fundamental Plane and Mass Plane}
\label{sec:fp}

\begin{figure*}   
    \centering  
    \includegraphics[width=.49\textwidth]{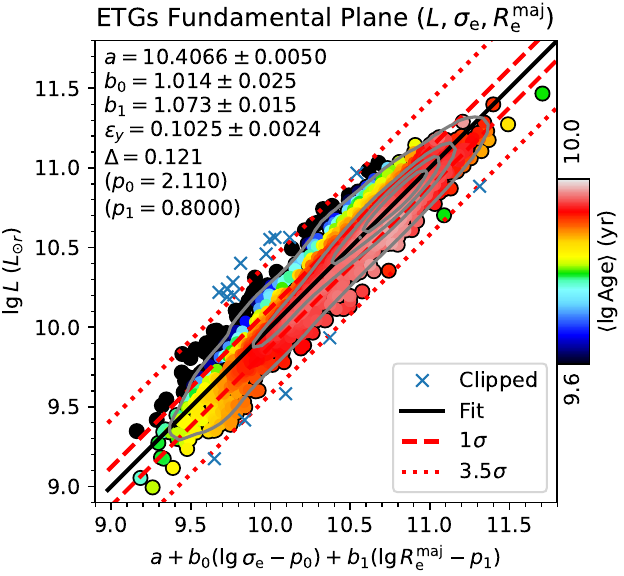}    
    \includegraphics[width=.49\textwidth]{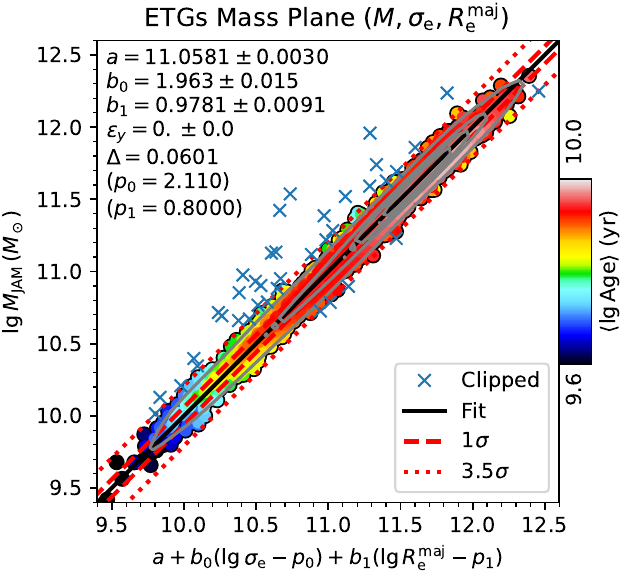}    
    \caption{
        Left panel: The Fundamental Plane of ETGs \citep[classified as $T_\mathrm{type}\le-0.5$ by][]{VazquezMata2022} for the MaNGA sample, comprising $N_\mathrm{ETG}=1868$ ETGs with $\text{Qual}\ge1$ from the DynPop sample of \citet{Zhu2023}. The plane shows significant scatter, primarily due to a strong age dependency orthogonal to the plane.        
        Right panel: The Mass Plane, obtained by transforming the luminosity from the left panel into masses as $M_\mathrm{JAM} = (M/L)_\mathrm{JAM} \times L$ using the mass-follow-light JAM\textsubscript{cyl} dynamical models fitted to MaNGA stellar kinematics. The coefficients of the mass plane are close to the virial predictions $b_0=2, b_1=1$, and the observed rms scatter reduces from $\Delta=0.12$ to $\Delta=0.06$ dex, with intrinsic scatter consistent with zero, and any age trend orthogonal to the plane disappears. The fit was performed using the robust \href{https://pypi.org/project/ltsfit/}{ltsfit} procedure by \citet{Cappellari2013p15}, and crosses represent the outliers automatically removed from the fit.    
        The filled disks represent ETGs, colored by the loess-smoothed global luminosity-weighted age (using \href{https://pypi.org/project/loess/}{loess.loess\_2d}; \citealt{Cappellari2013p20}). The gray contours show the kernel density estimate of the density distribution for the entire galaxy population of ETGs and spiral galaxies (using \href{https://docs.scipy.org/doc/scipy/reference/generated/scipy.stats.gaussian_kde.html}{scipy.stats.gaussian\_kde}; \citealt{Scipy2020}). This figure was adapted from \citet[figs.~3,4]{Zhu2024}.
        \label{fig:fp}}
    \end{figure*}

When the first kinematic measurements became possible, it was found that elliptical galaxies follow a relation $L\propto\sigma^4$ between their total luminosity and stellar velocity dispersion \citep{Faber1976}. Shortly after, a relation was discovered between the galaxy's average surface brightness $\Sigma_\mathrm{e}$ inside $1\re$ and their luminosity \citep{Kormendy1977}. 

To understand the meaning of these relations, one can use the virial theorem (\citealt[sec.~4.8]{Binney2008}; \citealt[sec.~V]{Courteau2014}). Galaxies may be expected to satisfy the virial equilibrium condition, which applies to isolated systems of particles moving under the influence of mutual gravity. For a distribution of stars in a steady state with spherical symmetry, the scalar virial relation is:
\begin{equation}\label{eq:scalar_virial}
    M = 3\frac{r_g\, \langle v_\mathrm{LOS}^2\rangle_\infty}{G},
\end{equation}
where $\sigma_\mathrm{LOS}^2\equiv\langle v_\mathrm{LOS}^2\rangle_\infty$ is the line-of-sight second velocity moment integrated over all particles, $r_g \equiv 2 M^2/\int_0^\infty [M(r)/r]^2 dr$ is the gravitational radius of the particle distribution, and $M$ is their total mass. The relation is \emph{rigorously} insensitive to the anisotropy or orbital distribution of the stars. The radius $r_g$ varies between 2.2 and 4.4 times the half-light radius $\re$, for an assumed \ser\ projected mass distribution with $n=1-10$ \citep[fig.~6]{Cappellari2013p15}.

A problem with \autoref{eq:scalar_virial} is that the relevant quantities are not directly observable in real galaxies: (i) the kinematics can only be measured out to a finite radius and (ii) galaxies are thought to contain both luminous and dark matter, which means that $M$ is not the mass of the visible stars from which we measure $\sigma$. This breaks the validity of the equation. For this reason, \autoref{eq:scalar_virial} can only be used for studying qualitative trends and not for quantitative measurements. Nonetheless, the fact that the newly discovered \citet{Faber1976} relation did not follow the virial prediction was interpreted as indicating a smooth variation of the mass-to-light ratio $M/L$ among galaxies. 

A decade later, astrophysicists realized that the two previous relations are just special projections of a more general relation, aptly named the Fundamental Plane \citep[FP,][]{Dressler1987, Djorgovski1987}, between $(L,\se,\Sigma_\mathrm{e})$. Currently, one defines $L$ as the galaxy's total luminosity; $\re$ is the radius enclosing half of the total luminosity; and $\se$ is the velocity dispersion integrated within a galaxy isophote of area $\pi\re^2$. Initially, astrophysicists used $\Sigma_\mathrm{e}$ as one of the three variables of the FP because both $\se$ and $\Sigma_\mathrm{e}$ are distance-independent (ignoring cosmological surface-brightness dimming). This allows the FP to be used for measuring galaxy distances by scaling $L$ to match a reference FP. However, $\Sigma_\mathrm{e}=L/(2\pi\re^2)$, which means that $L$ appears in two of the FP coordinates, creating strong correlations. Nowadays, we think we know the geometry of our Universe quite accurately, and redshift is generally a more reliable distance indicator than the FP for distant galaxies \citep[but see][]{DEugenio2024hyper}. To reduce the correlation between variables, a better form for the FP is $L\propto\se^b\re^c$ \citep{Cappellari2013p15}, which in logarithmic variables becomes the plane:
\begin{equation}
    \lg L = a + b \lg\se + c\lg\re.
\end{equation}

If galaxies were self-similar stellar systems with the same $M/L$, then virial equilibrium would imply a relation $M\propto L\propto\se^2\re$. However, it was clear from the first determinations that the FP exponents are very different from the virial prediction $b=2$, $c=1$. Numerous determinations for different samples and with different techniques have confirmed this so-called `tilt' of the FP over the past forty years \citep[e.g.,][]{Colless2001, Bernardi2003fp, Cappellari2013p15, DEugenio2021}. As a recent example, the determination for the largest sample of ETGs of the MaNGA sample using high-quality IFS kinematics gives $b=0.98\pm0.02$, $c=1.03\pm0.02$ \citep{Zhu2024}.

After the discovery of the tilt, a rather long debate ensued \citep[see review][sec.~4.1.2]{Cappellari2016}. This is because the tilt could be explained in several ways: (i) It could be a genuine variation in the $M/L$. This could be due to either changes in the stellar population ages or metallicity, or to changes in the dark matter fraction (\autoref{sec:total_density}). (ii) Alternatively, the tilt could be due to non-homology, namely the fact that galaxies change structural properties as a function of their mass or velocity dispersion. In particular, we have seen that galaxies vary both their photometric profiles (\autoref{sec:sersic_profiles}) and their kinematics (\autoref{sec:lam_eps}) as a function of stellar mass.

One way to distinguish between the two alternatives is to directly measure the $M/L$ of galaxies while modeling non-homology effects in detail. This can be done with either dynamical modeling \citep[sec.~3.4]{Cappellari2016} or strong gravitational lensing \citep{Treu2010} techniques. A first detailed study using \citet{Schwarzschild1979} dynamical modeling of a sample of 25 galaxies concluded unambiguously that the tilt of the FP was due almost entirely to a genuine $M/L$ variation, and more specifically driven by a tight $(M/L)-\se$ relation, with homology having a negligible contribution \citep{Cappellari2006}. This result was later confirmed with samples of ever-increasing size and using the JAM method. The study of \citet{Cappellari2013p15} modeled the 260 nearby ETGs of the ATLAS\textsuperscript{3D} sample; this was extended to about 2000 galaxies of all morphological types from the MaNGA sample by \citet{Li2018}, and to the full MaNGA galaxy sample of 10K galaxies by \citet{Zhu2024}. The same conclusion was reached using 36 strong gravitational lens ETGs by \citet{Bolton2007mp} and 73 lens ETGs by \citet{Auger2010}. All these studies transformed the FP into the Mass Plane (MP) by multiplying the luminosity by the total $M/L$ within the central regions (typically $1\re$). They consistently found that, unlike the FP, the MP essentially follows the Virial predictions. An illustration of this fact is shown in \autoref{fig:fp} from the currently largest MaNGA detailed dynamical modeling study by \citet{Zhu2024}. Importantly, the precise coefficients of the FP and MP were found to depend sensitively on how the parameters $(M,\se,\re)$ are measured (see the papers for details).

Given the rather small fraction of dark matter in the central region of ETGs (\citealt{Cappellari2013p15, Santucci2022, Zhu2024}; \autoref{fig:total_slope} right), the $M/L$ variation must be almost entirely driven by variations in the stellar population \citep{Graves2009b, FalconBarroso2011, Magoulas2012, DEugenio2021}, including variations in the stellar initial mass function \citep{vanDokkum2010, Cappellari2012, Cappellari2013p20, Spiniello2012}. Given that nearly all $M/L$ variation happens as a function of the stellar velocity dispersion, as I will discuss in \autoref{sec:mass-size}, the tilt of the FP can be accurately predicted by the $(M/L)-\se$ relation alone \citep[sec.~4.2]{Cappellari2016}.

\subsection{The Virial Mass Estimator}
\label{sec:virial}

For maximum accuracy in determining the dynamical masses of galaxies, one should use dynamical models like JAM or Schwarzschild's method. JAM is more accurate when only $V$ and $\sigma$ can be reliably measured from the single-aperture of integral-field data. Schwarzschild's method can exploit the highest-quality integral-field kinematics, fitting the higher-order moments of the LOSVD. Both models require minimal assumptions and can model the instrumental PSF, the size of the observed aperture, or fit two-dimensional kinematics in detail. Nonetheless, the virial estimator has been used in numerous papers and holds historical significance. It is still used today in some cases.

The mass of a galaxy is generally an ill-defined concept from an empirical standpoint. Galaxies are extended objects with fuzzy boundaries, consisting of stars in the central regions and extended dark halos further out. Except for very rare situations, there are no tracers, either from dynamics or gravitational lensing, that can sample the full extent of an individual galaxy and allow us to measure its total mass. Only weak gravitational lensing can come close to measuring the galaxy's total mass, but this is done in an assumption-dependent statistical manner, rather than for individual objects \citep[see review by][]{Mandelbaum2018}. Instead, the quantity that both dynamical models and strong gravitational lensing directly measure is the density profile within the region covered by the tracer (the FoV of the kinematic observations, or the Einstein radius of the lens, which is generally $<\re$). The variation in density is dominated by variations in surface brightness, which varies by many orders of magnitude within a galaxy and is therefore not very meaningful. Much more meaningful is the $M/L$ within the region covered by the tracer, which generally varies spatially by less than a factor of 1.5. 

Two common ways to measure the virial parameters $(L,\re)$ are: (i) by fitting \ser\ profiles to the galaxy images, or (ii) essentially non-parametrically, using the Multi-Gaussian Expansion \citep{Emsellem1994, Cappellari2002mge}. Using \ser\ profiles is especially common at high redshift \citep[e.g.,][]{vanderWel2014}, where the spatial resolution is limited, and the \ser\ profile captures most of the information in the data, while MGE is often used with well-resolved photometry \citep[e.g.,][]{Cappellari2013p15}. The \ser\ profile is also useful to study correlations with other galaxy properties (e.g., \autoref{fig:sersic}).

In either case, one can construct a version of the scalar virial relation from \autoref{eq:scalar_virial} for quantitative determinations as follows: (i) measure the well-defined total $M/L$ within a given spherical radius $r_M$, instead of the poorly-defined total mass; (ii) measure the luminosity-weighted $\sigma_\mathrm{LOS}$ within a finite aperture $R_\sigma$. A common choice of radii is $r_M=R_\sigma=\re$, in which case $\se^2\equiv \langle v_\mathrm{LOS}^2\rangle_{\re}$ and the virial relation becomes:
\begin{equation} 
    (M/L)(r<\re)\approx k_\textrm{vir}\, \frac{\re\se^2}{G L},
\end{equation}
where $k_\textrm{vir}$ is a coefficient calibrated using dynamical models. It was found that $k_\textrm{vir}$ crucially depends on how the virial parameters are measured \citep[fig.~14]{Cappellari2013p15}, as well as on the galaxies' inclination \citep{vanderWel2022}:
\begin{description}
    \item[MGEs:] When $(\re,L)$ are measured using MGEs, which do not involve extrapolating the galaxy luminosity $L$ to infinite radii, the coefficient $k_\textrm{vir}$ is nearly constant around $k_\textrm{vir}=4-5$ \citep{Cappellari2006}. However, the precise value of $k_\textrm{vir}$ depends on the depth of the observed photometry, making this method ideal for obtaining accurate scaling relation with homogeneous data, but generally unreliable for accurate absolute $M/L$ determinations;
    \item[\ser:] When $(\re,L)$ are measured by fitting \ser\ photometric models, and $L$ represents the total galaxy luminosity extrapolated to infinite radii, then $k_\textrm{vir}(n)$ is a strong function of the \ser\ index. The functional form of $k_\textrm{vir}$ can be predicted using models in absolute terms. However, this method relies on the assumption that galaxies under study are accurately described by the \ser\ profile, which is generally a crude approximation.
\end{description}

A common expression for $k_\textrm{vir}(n)$ can be found in \citet[eq.~20]{Cappellari2006}. However, this expression was computed for the interval $n=[2,10]$. As such, it misses the maximum of $k_\textrm{vir}(n)$ at $n\approx1.1$ and becomes significantly inaccurate for smaller values of $n$ (e.g., 19\% error at $n=0.5$), which are typical in spiral galaxies (see \autoref{fig:sersic}, right panel). To address this, I utilized the \href{https://pypi.org/project/jampy/}{\texttt{jam.sph.proj}} function within the JAM package\footnote{\url{https://pypi.org/project/jampy/}} to compute the virial coefficient for various $\lg n$ values. This computation assumed the same spherical isotropic condition and integrated within an aperture of radius $R=\re$ as done by \citet{Cappellari2006}. I then derived a rational-function approximation for $k_\textrm{vir}(n)$, with a maximum relative error of 0.4\% over an extended interval $n=[0.5,16]$.
\begin{equation}
    \lg k_\textrm{vir}(n) = \frac{0.8794 - 0.3405 \lg n - 0.2636 (\lg n)^2}{1 - 0.4194 \lg n}.
\end{equation}
In the interval $n=[2,10]$ this function agrees within 0.5\% with the previous one as expected.

\subsection{The Mass Versus Size Distribution}
\label{sec:mass-size}

\begin{figure*}    
    \centering  
    \includegraphics[width=\textwidth]{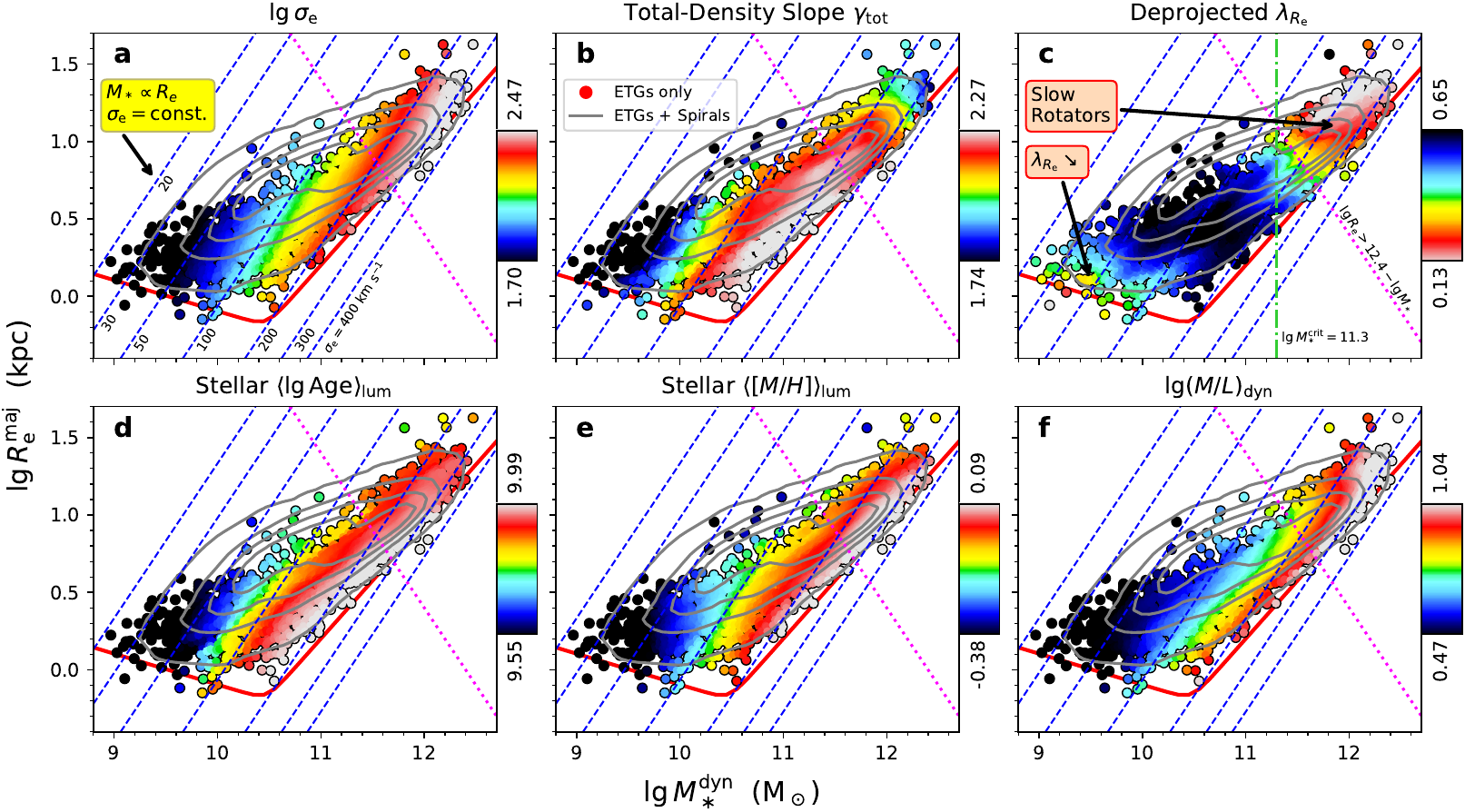}    
    \caption{Distributions of global physical properties of 2729 early-type galaxies (ETGs) on the stellar mass–effective radius ($M_*$, $\re^{\rm maj}$) plane, displaying six parameters: (a) effective luminosity-weighted velocity dispersion $\se^2 \equiv \langle V^2 + \sigma^2\rangle$ within the half-light radius $\re$; (b) mass-weighted total density slope (see \autoref{eq:slope}) from cylindrically-aligned JAM\textsubscript{cyl} models with NFW dark halos; (c) deprojected intrinsic specific angular momentum $\lam^\mathrm{intr} = \lam/\sin i$, edge-on corrected using JAM\textsubscript{cyl}-derived inclinations $i$; (d) luminosity-weighted stellar age; (e) luminosity-weighted metallicity; and (f) total dynamical mass-to-light ratio $M/L$ from mass-follows-light JAM\textsubscript{cyl} models. Distributions are smoothed via the LOESS algorithm (\href{https://pypi.org/project/loess/}{loess.loess\_2d}; \citealt{Cappellari2013p20}) with \texttt{frac = 0.1}. Dashed curves trace lines of constant effective velocity dispersion $\se$ from the scalar virial relation $\se^2 \equiv GM_{\rm JAM}/(5 \re^{\rm maj})$ (factor 5 from \citealt{Cappellari2006}). The red solid curve marks the zone of exclusion (ZOE) from \citet{Cappellari2013p20}, while grey contours show the kernel density estimate of galaxy density (including ETGs and spirals) using \href{https://docs.scipy.org/doc/scipy/reference/generated/scipy.stats.gaussian_kde.html}{scipy.stats.gaussian\_kde} (\citealt{Scipy2020}). In all panels, excluding (c), trends follow constant-$\se$ lines (i.e., $M_* \propto \re$), below the high-mass slow-rotator boundary defined by the magenta dotted line $\lg(\re/\text{kpc}) \gtrsim 12.4 - \lg(M_*/M_\odot)$. Panel (d) diverges: trends align orthogonally to constant-$\se$ lines. Massive slow rotators dominate above $\lg(M_*^\mathrm{crit}/M_\odot) \gtrsim 11.3$ (or more precisely above the dotted magenta boundary), while a distinct low-mass population of slow-rotating galaxies (arrow-indicated) emerges below $\lg(M_*/M_\odot) \lesssim 9.5$, though their rotation is less suppressed compared to the extreme slow rotation of their massive counterparts. Parameters are sourced from the MaNGA DynPop catalog: dynamical quantities (catalog keywords \texttt{Sigma\_Re}, \texttt{MW\_Gt\_Re}, \texttt{Lambda\_Re}, \texttt{log\_ML\_dyn}, \texttt{Lum\_tot\_MGE}, \texttt{Rmaj\_kpc\_MGE}) from \citet{Zhu2023}, and stellar population properties (catalog keywords \texttt{LW\_Age\_Re}, \texttt{LW\_Metal\_Re}) from \citet{Lu2023}. Panels (a)--(c) were originally presented for the combined spiral and ETG sample in \citet[fig.~16]{Zhu2024}, while panels (d)--(f) in \citet[fig.~8]{Lu2023}.  
  \label{fig:mass-size}}  
\end{figure*} 

\begin{figure*}   
    \centering  
    \includegraphics[width=.7\textwidth]{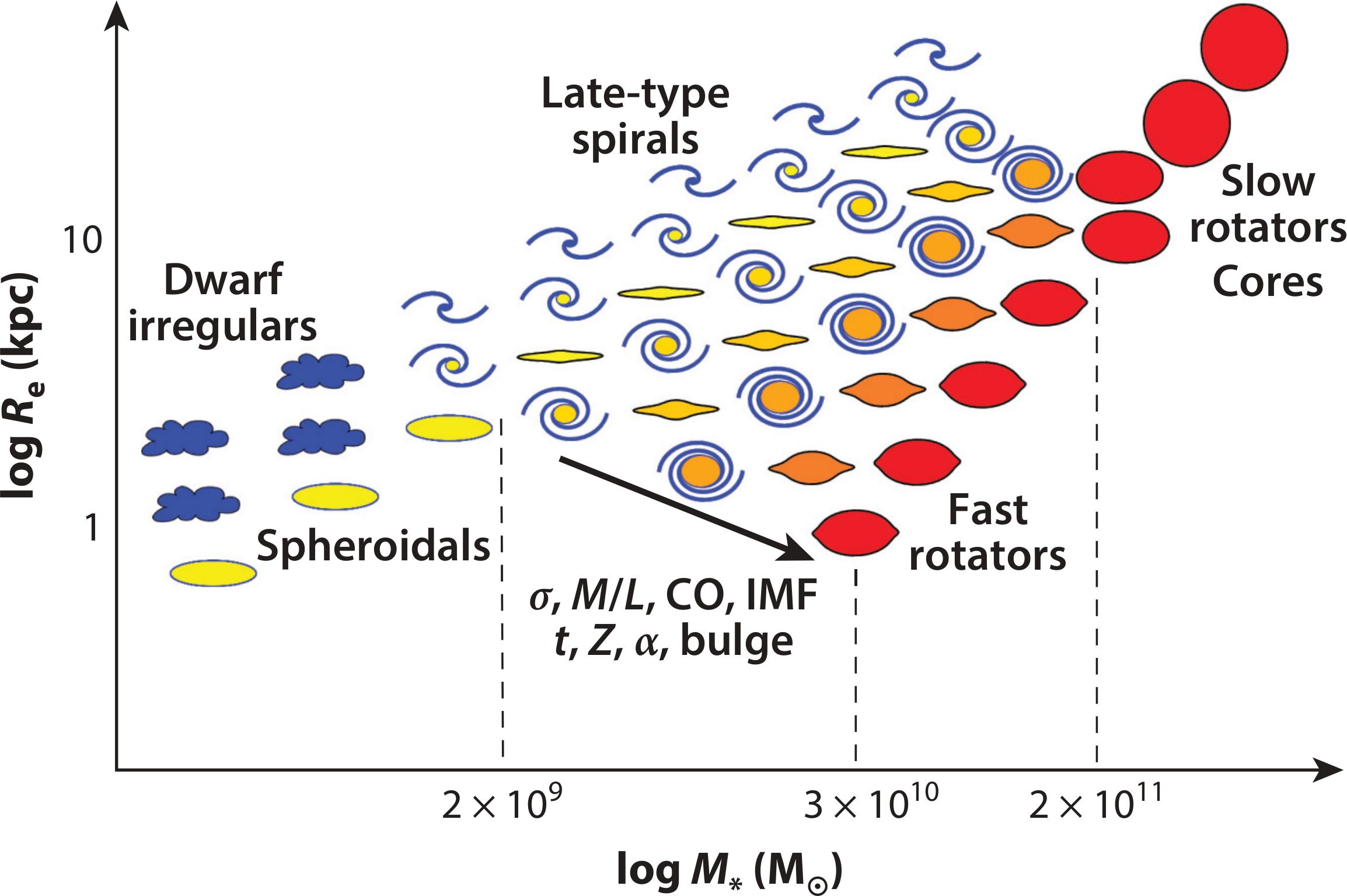}    
    \caption{Schematic distribution of galaxy properties on the stellar mass ($M_*$) vs. half-light radius ($\re$) plane. Early-type galaxy (ETG) characteristics—such as stellar population parameters (ages, metallicity, $\alpha$-enhancement, IMF) and gas content (CO mass fraction)—correlate along lines of nearly constant effective velocity dispersion $\se$ (equivalent to $\re \propto M_*$), tracing the bulge mass fraction or the steepness of the total mass density profile ($\gamma_\mathrm{tot}$). This sequence transitions smoothly into the spiral galaxy population: minimal overlap occurs between late-type spirals (Sc-Irr) and ETGs, significant overlap exists between early-type spirals (Sa-Sb) and low mass-to-light ratio ($M/L$) fast rotators, and no overlap is seen between spirals and high-$M/L$ fast rotators. Three key mass scales define distinct regimes:  
    (i) $M_* \lesssim 2 \times 10^9 \, M_\odot$: Regular ETGs are absent, and the lower boundary of the mass-size relation rises with increasing mass.  
    (ii) $M_* \approx 3 \times 10^{10} \, M_\odot$: ETGs reach minimum sizes (or maximum stellar densities), with the slope of the mass-size  lower boundary steepening to $\re \propto M^{0.75}$ at higher masses.  
    (iii) $M_* \lesssim 2 \times 10^{11} \, M_\odot$: ETGs are dominated by flat fast rotators with disks. Above this threshold, spirals become rare, and the population transitions to round/weakly triaxial slow rotators with flat (core/deficit) central surface brightness profiles. \citep[adapted from][fig.~14]{Cappellari2013p20}  
    \label{fig:mass_size_cappellari2016}}  
\end{figure*}

\begin{figure*}    
    \centering   
    \includegraphics[width=.49\textwidth]{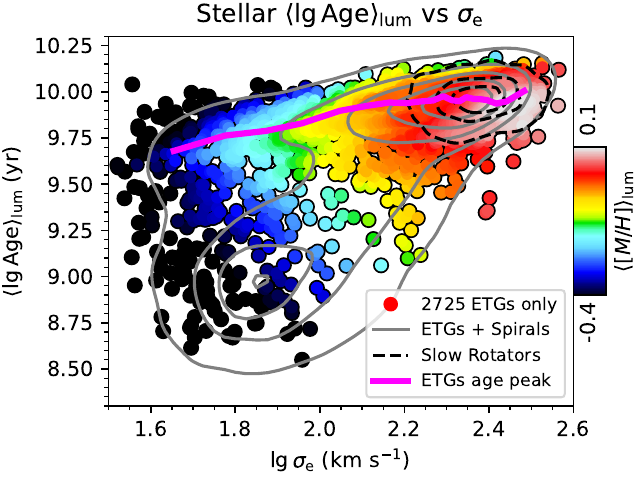}    
    \includegraphics[width=.49\textwidth]{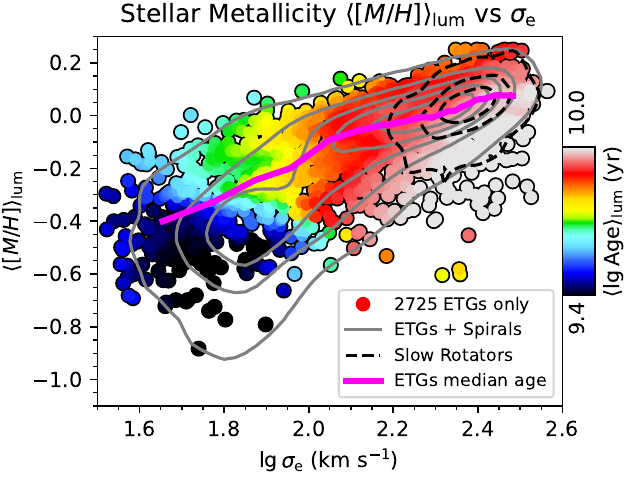}    
    \caption{Left panel: Distribution of global luminosity-weighted mean age $\langle \lg \text{Age} \rangle_{\mathrm{lum}}$ within the effective radius $\re$ versus effective velocity dispersion $\se$ for early-type galaxies (ETGs; $T_{\mathrm{type}} \leq -0.5$; \citealt{VazquezMata2022}). Symbols are colored by the LOESS-smoothed global luminosity-weighted metallicity $\langle [M/H] \rangle_{\mathrm{lum}}$ within $\re$ (via \href{https://pypi.org/project/loess/}{loess.loess\_2d}; \citealt{Cappellari2013p20}). The magenta solid line traces the mode of the age distribution at fixed $\se$, calculated in 20 equal-number $\se$ bins by identifying the peak of a kernel density estimate (KDE) in each bin. This method emphasizes the narrow ridge in the asymmetric age distribution. Right panel: Distribution of global luminosity-weighted metallicity $\langle [M/H] \rangle_{\mathrm{lum}}$ versus $\se$, colored by $\langle \lg \text{Age} \rangle_{\mathrm{lum}}$; the magenta line shows the median metallicity in $\se$ bins. In both panels, gray contours represent the KDE of the density distribution for all ETGs and spiral galaxies (using \href{https://docs.scipy.org/doc/scipy/reference/generated/scipy.stats.gaussian_kde.html}{scipy.stats.gaussian\_kde}; \citealt{Scipy2020}), while black dashed contours correspond to slow rotators. Data are from the MaNGA DynPop catalog: $\se$ values (\texttt{Sigma\_Re}) from \citet{Zhu2023}, and stellar ages/metallicities (\texttt{LW\_Age\_Re}, \texttt{LW\_Metal\_Re}) from \citet{Lu2023}. A combined version (ETGs + spirals) appears in \citet[Fig.~6]{Lu2023}. \label{fig:age-metal-sigma}}
\end{figure*}

\begin{figure*}   
    \centering  
    \includegraphics[width=.7\textwidth]{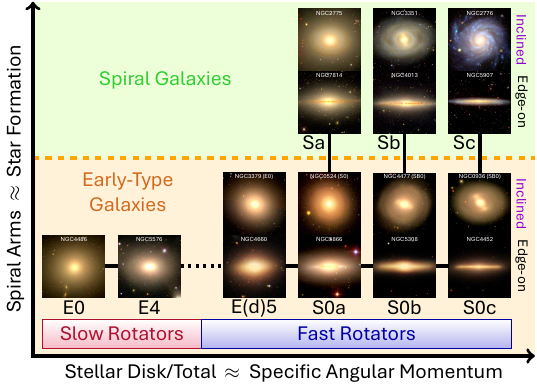}    
    \caption{This `comb' diagram maps the systematic relationships between morphological features (e.g., disk-to-total stellar mass fraction, spiral arm prominence) and stellar kinematic properties (e.g., angular momentum, rotation curves) of galaxies. The x-axis quantifies the disk-to-total mass ratio ($D/T$), which correlates with specific stellar angular momentum $\lam$ within the half-light radius $\re$ and disk spatial extent, while the y-axis reflects spiral arm strength, tied to star formation activity or stellar population age. Fast-rotator early-type galaxies (ETGs), when edge-on, resemble S0s or disky ellipticals \citep[E(d),][]{Kormendy1996}, while many show bars when at low inclination (e.g., NGC~4477 or NGC~0936), or dusty disks (e.g., NGC~0524 or NGC~5866), yet photometry alone often misclassifies them: $\sim$2/3 of morphologically labeled ellipticals (e.g., NGC~3379) are fast rotators. These systems span the full $D/T$ range: from spheroid-dominated galaxies with compact stellar disks confined to $\lesssim\re$ (e.g., NGC~4660, where the spheroid dominates at large radii) to galaxies dominated by thin, extended stellar disks prominent at large radii (e.g., NGC~4452). Kinematically, spheroid-dominated fast rotators display a stellar mean rotation curve that peaks within $\re$ and declines beyond it \citep[][fig.~3a]{Cappellari2016}, while galaxies with higher disk-to-total mass ratios ($D/T$) exhibit rotation peaks at larger radii \citep[][fig.~3b]{Cappellari2016} or reduced $\se$ at fixed stellar mass. Slow rotators are intrinsically round ($\eps^\mathrm{intr} \lesssim 0.4$) and always appear so in projection. Dynamical models and statistical inversion confirm fast rotators are flatter ($\eps^\mathrm{intr} \gtrsim 0.4$) than slow rotators. Solid lines denote empirical continuity; dashed lines mark the fast/slow rotator bimodality \citep[][fig.~1]{Cappellari2011b}. Flat ETGs are rare compared to flat spirals; diagram proportions do not reflect galaxy abundances. \label{fig:atlas3d_comb}}  
\end{figure*}

In \autoref{sec:fp}, I showed that the existence of the Fundamental Plane and Mass Plane are almost entirely due to virial equilibrium combined with a smooth variation of the galaxies' stellar population and $M/L$. Virial equilibrium applies to any isolated system, regardless of how it reached that equilibrium, implying limited information on galaxy evolution is encoded in those relations.

When we look at the FP or MP far from an edge-on view, we see that galaxies are not distributed randomly on that plane. Given the thinness of the MP, the perspective does not significantly affect the plane's appearance, provided we are not too close to an edge-on view. A particularly useful view of the MP is the $(M,\re)$ distribution (\autoref{fig:mass-size}), as both mass and size are quantities that are easy to measure and nearly uncorrelated observationally. Moreover, galaxy masses and sizes are expected to vary in a clear and qualitatively predictable way during galaxy evolution. For this reason, the $(M,\re)$ distribution represents one of the cleanest benchmarks against which to compare galaxy evolution models.

Here, the mass $M=M_*^\mathrm{dyn}$ represents the most accurate estimator of the total stellar mass $M_*$ \citep[see][]{Cappellari2013p15}. This is because dark matter is generally negligible in ETGs. Thus, the dynamical $(M/L)(r<\re)$ approximates the stellar population $(M_*/L)_\mathrm{pop}$, implying that $(M/L)(r<\re)\times L\approx (M_*/L)_\mathrm{pop}\times L\approx M_*$.

The clearest and most striking feature of \autoref{fig:mass-size} is that neither the galaxies' stellar mass nor sizes are the main drivers of galaxy properties. Instead, both the parameters related to galaxy dynamics, such as $M/L$ and dark matter fraction $f_\mathrm{DM}$, and parameters describing the stellar population, such as age and metallicity, are nearly constant along lines with $M_*\propto\re$, which correspond to lines of constant velocity dispersion $\se\propto\sqrt{M_*/\re}$ according to the virial estimator.

Only the deprojected stellar specific angular momentum $\lam$ behaves differently. Below $\mcrit\approx2\times10^{11}\,\msun$, there is no trends with $\se$. Nearly all ETGs in this low-mass regime have the large $\lam$ characteristic of fast rotators. These ETGs are a population of randomly inclined, nearly axisymmetric galaxies with stellar disks. However, above $\mcrit$ there is a sharp transition to much lower $\lam$ \citep{Emsellem2011,Cappellari2013p20,Veale2017,Graham2018}. This is the $M_*$ regime where slow rotators start appearing in the ETG population \citep[see review in][]{Cappellari2016}. Panel (c) of \autoref{fig:mass-size} reveals that and even better fast/slow rotator ETGs separation can be achieving by using a line orthogonal (i.e., with $\re\propto-M_*$) to lines of constant $\se$ (i.e., with $\re\propto M_*$) and with stellar mass above $M_*^\mathrm{crit}$
\begin{equation}\label{eq:mass_re_fast_slow_separation}
    \lg(\re/\text{kpc}) \gtrsim 12.4 - \lg (M_*/M_\odot).
\end{equation}

Below the limit defined by \autoref{eq:mass_re_fast_slow_separation} the trend in ETGs merges smoothly with that of spiral galaxies, which occupy the region of low $\se$ in the $(M,\re)$ plane \citep{Cappellari2013p20}. The population and dynamical properties of spiral galaxies also follow the same constant-$\se$ distribution as ETGs \citep{Li2018,Lu2023}. This parallelism is crucial for understanding ETG evolution. A similar parallelism between spiral galaxies and fast rotator ETGs can be observed when studying the star formation rate versus stellar mass $\text{SFR}-M_*$ relation \citep[e.g.,][]{Brinchmann2004,Noeske2007}. Below $\mcrit$, fast rotator ETGs lie as expected below the star-forming main sequence, which is populated by spiral galaxies, but show only a small increase in their specific angular momentum. This is due to their more massive bulges, compared to spiral galaxies of similar mass. In contrast, above $\mcrit$, in the region where slow rotators start dominating the ETGs population, a decrease in SFR is accompanied by a sharp decrease in $\lam$ \citep{Wang2020,Cortese2022}.

As seen from \autoref{eq:scalar_virial}, when the ratio of $M$ to $\re$ is constant, the integrated second velocity moment $\sigma^2_\mathrm{LOS}$ is expected to be nearly constant as well. Empirically, using masses from detailed dynamical models, it was found that the virial formula $\se^2=G M/(k_\textrm{vir}\re)$ accurately predicts the corresponding observable stellar $\se$ \citep{Cappellari2013p15}. The fact that galaxy properties follow lines with $M\propto\re$ indicates that the stellar velocity dispersion $\se$, or a virial estimate based on masses and sizes, better predicts galaxy properties than masses $M$, radii $\re$, or effective surface density $\Sigma_\mathrm{e}$. This correlation was initially pointed out from a small galaxy sample with dynamical models in the SAURON survey \citep{Cappellari2006}. It was confirmed and generalized to other stellar population parameters with dynamical models of ever-increasing samples of galaxies from the ATLAS\textsuperscript{3D} \citep{Cappellari2013p20, McDermid2015}, SAMI \citep{Scott2017, Barone2018}, and MaNGA surveys \citep{Li2018, Zhu2024}. Similar results were also found using the virial estimator alone in the SDSS survey \citep{Graves2009b} and using photometry alone \citep{Franx2008, Bell2012}.

A similarly good predictor of galaxy properties is the core density $\Sigma_1$, measured within a fixed aperture of $R=1$ pc, which is generally smaller than $\re$ and samples the central region of galaxies \citep{Cheung2012, Barro2017}. In some of the discussions that follow, I will focus on $\se$, but the same conclusions apply if one replaces the stellar velocity dispersion $\se$ with its virial estimate $\se\propto\sqrt{M/\re}$ or the core density $\Sigma_1$. The similarity between $\Sigma_1$ and $\se$ is illustrated in \citet{Fang2013, Cappellari2023}. The relation between galaxy properties and $\se$, leading to parallel sequences of equal $\se$, or equivalently $M_*\propto\re$, on the $(M_*,\re)$ diagram, was highlighted in the review by \citet[fig.~23]{Cappellari2016}, and I reproduce it here in \autoref{fig:mass_size_cappellari2016}.

The dependence of galaxy properties on $\se$ has important implications for understanding galaxy formation. At fixed stellar mass, both the velocity dispersion $\se$ and the core density $\Sigma_1$ quantify the concentration of stellar mass near the center, or the prominence of the stellar bulge or spheroid. The $\se$ is also the empirical quantity most tightly correlated with the mass of the central supermassive black hole (SMBH) \citep[see review by][]{Kormendy2013review}. The empirical correlation tells us that when galaxies quench their star formation and become passive and metal-rich ETGs, they also grow their bulges and central SMBHs.

However, it is essential to note that the dependence of the stellar population on $\se$ is not smooth and monotonic. There is a critical value of $\lg(\scrit/\kms)\approx2.3$ \citep{Cappellari2023} or $\Sigma_1$ \citep{Barro2017, Chen2020}, above which all galaxies are quenched—namely, non-star-forming, old, and metal-rich. For lower $\se$ values, there is a large range and a bimodal distribution of population parameters. In particular, there is a clear trend with metallicity $[Z/H]$ being lower for younger galaxies at a given $\se$. This is illustrated in \autoref{fig:age-metal-sigma} for the final sample of 10K galaxies of the MaNGA survey \citep{Lu2023}.

The $(\mathrm{Age},\se)$ diagram is closely related to the color-magnitude diagram \citep{Faber2007, Schawinski2014}, or the $(M_*,\mathrm{SFR})$ diagram \citep{Brinchmann2004,Noeske2007} between stellar mass and star-formation rate (SFR). These diagrams are often used to understand the process of galaxy quenching. The advantage of the $(\mathrm{Age},\se)$ diagram is that $\se$, unlike mass $M_*$ or luminosity $L$, better aligns with the main direction along which galaxy properties vary. This reduces the overlap between different star formation histories of galaxies with different $M_*$. In this diagram, ETGs are located mainly in the old sequence, at the top of the diagram, equivalent to the red sequence of the color-magnitude diagram. However, below $\lg(\se/\kms)\approx2.3$, younger ETGs start appearing, having on average lower metallicity and straddling the region populated by spiral galaxies, which tend to lie at the lowest $\se$ and lowest metallicities.

The parallelism between the properties of fast rotator ETGs and spiral galaxies in the $(M,\re)$ plane led \citet{Cappellari2011b} to propose a revision to the classic tuning fork \citep{Hubble1936} of \autoref{fig:tuning_fork}. In this `comb diagram' for galaxy classification, fast rotator ETGs form a parallel sequence to spiral galaxies (\autoref{fig:atlas3d_comb}). Fast rotator ETGs share the same range of disk fractions (S0a--S0c) as spiral galaxies but also extend towards more extreme spheroid-dominated disky-elliptical galaxies E(d)5 \citep{Kormendy1996}. The nomenclature S0a--S0c for edge-on fast rotators was taken from \citet{vandenBergh1976}, who previously suggested a similar parallelism between the subset of S0 alone and spiral galaxies. The difference between the `trident' diagram for E, S0s, and spirals by \citet{vandenBergh1976} and the `comb' diagram for fast/slow rotator ETGs and spirals by \citet{Cappellari2011b} is that, the latter classification is based on kinematics, not photometry: fast rotators also include about $2/3$ of the inclined disk galaxies that have been misclassified morphologically as elliptical from photometry alone. These can be recognized at any inclination with stellar kinematics and constitute a much larger class of galaxies than S0s alone. Moreover, the `trident' did not include the extension to E(d) galaxies, which were not known at that time, and can only be detected from photometry when very close to edge-on.

The `comb' diagram in \autoref{fig:atlas3d_comb} underscores the significance of the variation in the disk-to-total ratio ($D/T$), and relative disk scale length, within the fast rotator population. On the right side of the `comb' handle are the nearly bulgeless S0c pure-disks fast rotators, resembling thin spiral galaxies stripped of their gas. In the middle are the S0b fast rotators, representing Hubble's typical S0 galaxies, with a central bulge and a disk dominating at larger radii. The left side of the fast rotator $D/T$ sequence is occupied by the spheroid-dominated fast rotators, which would be classified as disky-elliptical E(d) when viewed edge-on. These galaxies are characterized by stellar disks that dominate the central regions ($R\lesssim\re$) rather than the outer parts, and spheroids dominating at large radii.

The variation in $D/T$ and the shift from a disk dominating the outer parts to one dominating the central regions can be observed in kinematics at nearly every inclination. Specifically, when the disk dominates at large radii, stellar rotation continues increasing at $R \gtrsim \re$, whereas when the spheroid dominates at large radii, kinematics show a clear drop at $R \gtrsim \re$ \citep[see][fig.~3]{Cappellari2016}. This sequence of $D/T$ in ETGs is illustrated by \citet[fig.~7]{Graham2016}, who retain the same (but rotated) sequence as the ETGs comb handle of \autoref{fig:atlas3d_comb}, while slightly rearranging spiral galaxies. They propose the term `ellicular' (ES) for what I described as disky-elliptical E(d) fast rotators, characterized by centrally-concentrated disks. A detailed overview of galaxy classification schemes was presented by \citet{Graham2019ellicular}.

\section{Environmental Dependency of ETGs Properties}

\begin{figure*}   
    \centering  
    \includegraphics[height=.33\textwidth]{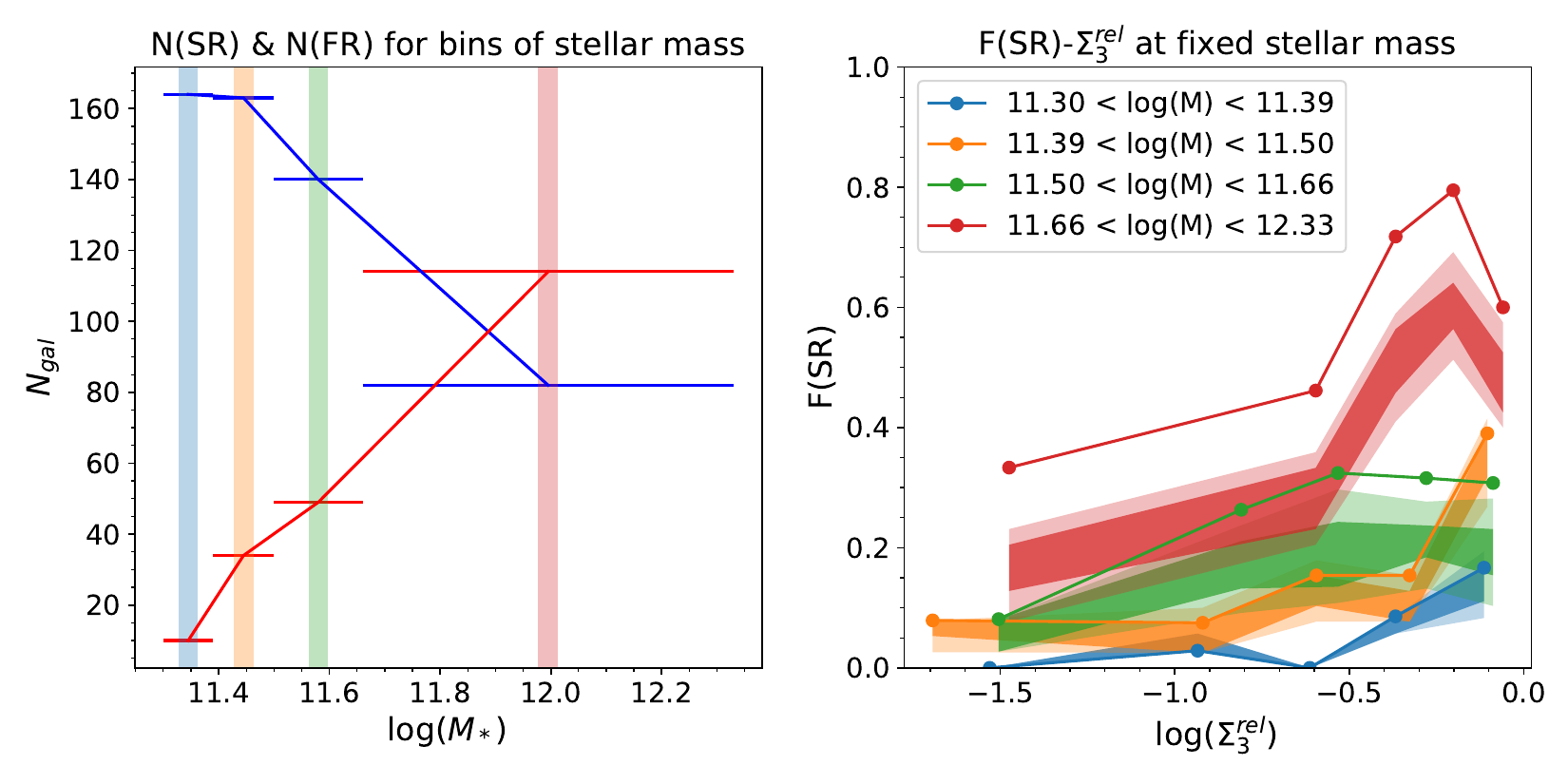}    
    \includegraphics[height=.33\textwidth]{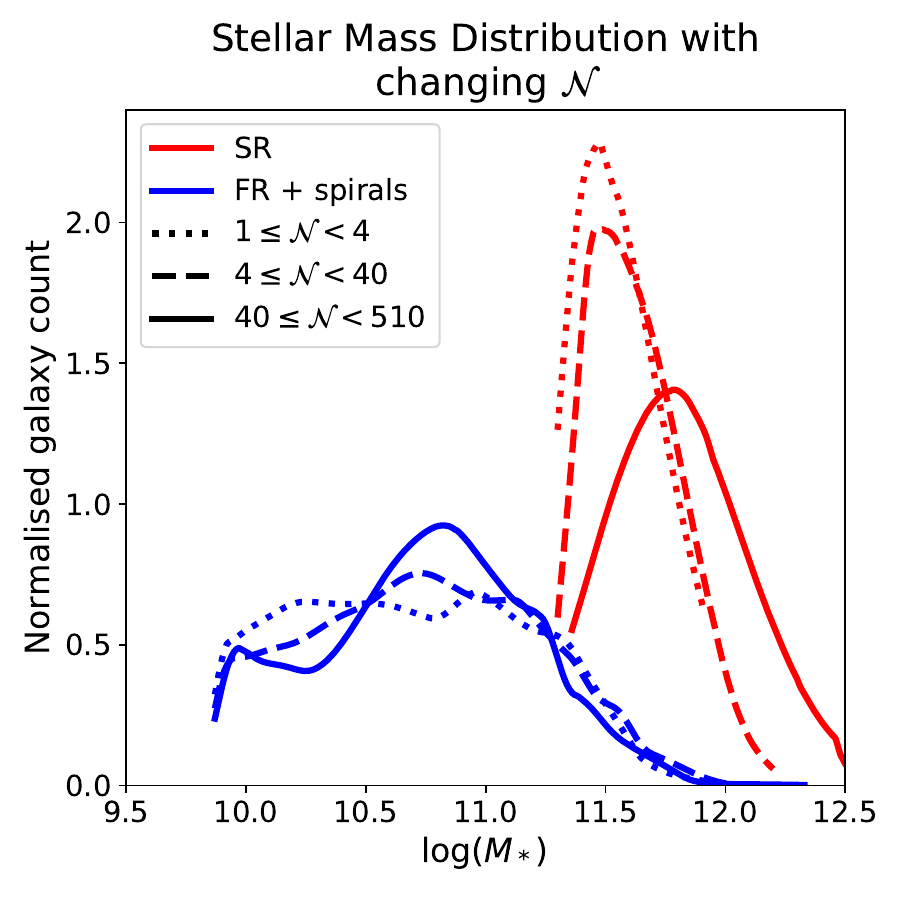}    
    \caption{ 
    Left panel: Fraction of fast vs. slow rotators (SRs) among galaxies above the critical mass $\lg(\mcrit/\msun) = 11.3$, as a function of stellar mass. Bins (each containing $\sim$200 ETGs) show SR fractions are negligible below $\mcrit$ but rise sharply above it, dominating at $\lg(M_*/\msun) \gtrsim 12$.  
    Middle panel: For bins in the left panel, SR fraction $F(\mathrm{SR})$ vs. galaxy number density $\Sigma_3^\mathrm{rel}$ (relative to cluster peak density). Shaded bands account for measurement uncertainties in $\lam$ and $\eps$, addressing misclassification bias. At fixed mass, the most massive SRs preferentially inhabit densest cluster regions.  
    Right panel: Stellar mass probability distributions for massive SRs (red) and non-SR galaxies (blue), split by environment: field (dotted), groups (dashed), clusters (solid). Curves are truncated below the completeness limit. While spirals/fast rotators show no environmental mass dependence, SRs shift to higher masses in denser environments \citep[fig.~7, 8]{Graham2019_environment}.  
    \label{fig:kinematics_fixed_mass}}      
\end{figure*}

\begin{figure*}   
    \centering  
    \includegraphics[width=.8\textwidth]{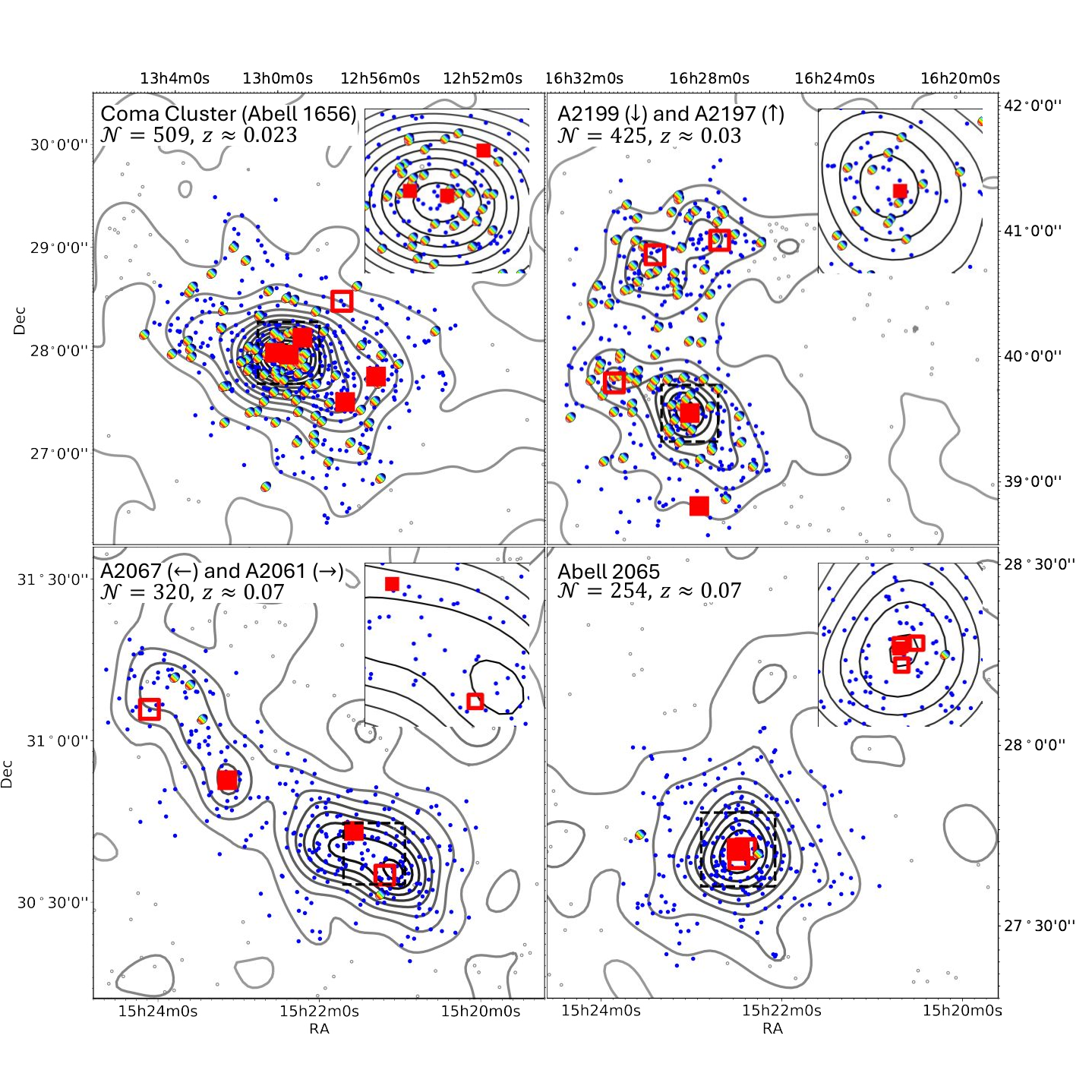}    
    \caption{  
        Galaxy clusters with MaNGA kinematics. Cluster members: red squares = slow rotators (SRs) with $M_* > \mcrit=2\times10^{11}\msun$; blue circles = other galaxies. Filled symbols denote FR/SR classifications from integral-field kinematics; unfilled symbols indicate photometric classifications. Gray unfilled circles are non-members. Contours show the kernel density estimate (KDE) of galaxy number density. Each panel (1 Mpc$^2$ field of view, centered on the densest peak; inset zooms in on core) lists the Abell designation, member count ($N$), and redshift ($z$). SRs are rare and preferentially located near density peaks, particularly in unrelaxed clusters with multiple subpeaks (each often hosting an SR).  
        Top Left: Coma Cluster (A1656) with three SRs at its core.  
        Top Right: A2197 and A2199. A2197's bimodal velocity distribution suggests two interacting subclusters: A2197W and A2197E.  
        Bottom Left: A2061 (lower right) and A2067 (upper left) in the Corona Borealis Supercluster (CBS), likely gravitationally bound.  
        Bottom Right: A2065 (CBS).  
        \citep[fig.~1]{Graham2019_clusters}  
        \label{fig:slow_rotator_in_clusters}}          
\end{figure*}

Understanding galaxy evolution fundamentally depends on the environment in which galaxies currently exist. Galaxies evolve hierarchically, with their present characteristics influenced by mergers and accretion from their surroundings. Consequently, the environment—typically quantified by the number density of galaxies around a given galaxy down to a certain mass limit—has long been correlated with galaxy properties.

The morphology-density relation is a cornerstone in the study of galaxy evolution, first brought to light by \citet{Dressler1980}. This relation reveals a profound connection between the density of a galaxy's environment and its morphological type. Dressler's extensive study, which involved observations of over 6000 galaxies across 55 clusters, highlighted a striking pattern: dense environments, such as the crowded centers of galaxy clusters, predominantly host early-type galaxies, including ellipticals and S0s. In stark contrast, spiral galaxies are more frequently found in less dense, more isolated regions.

This discovery underscored the significant influence that a galaxy's surroundings have on its evolutionary path. In high-density regions, various processes such as ram-pressure stripping, where the interstellar medium is stripped away by interaction with the intracluster medium, and galaxy harassment, involving frequent high-speed encounters between galaxies, contribute to transforming spiral galaxies into ETGs. These mechanisms help explain the scarcity of spiral galaxies in cluster cores.

The implications of the morphology-density relation are far-reaching. It suggests that the environment a galaxy resides in can fundamentally alter its development, affecting not only its structure but also its star formation history, color, and other properties. This has led to a broader understanding that galaxy evolution is a complex interplay between intrinsic properties and external influences.

Despite the many advancements in observational technology and data analysis since 1980, Dressler's findings have remained remarkably robust. Modern surveys, such as the Sloan Digital Sky Survey (SDSS) with the large Galaxy Zoo sample of morphological classifications of $10^5$ galaxies \citep[e.g.,][]{Bamford2009, Skibba2009}, have confirmed the morphology-density relation across different environments. Studies have also shown that the relation extends to redshifts $z\sim1$ \citep[e.g.,][]{Stanford1998, vanDokkum2000}. JWST is currently pushing the frontier, with some initial indications that environmental effects were already at work up to $z\sim6$, within one billion years after the Big Bang \citep[e.g.,][]{Morishita2024}. This consistency over time suggests that the environment has long played a crucial role in shaping galaxies.

The discovery that the vast majority of elliptical galaxies are actually misclassified S0-like galaxies seen at low inclination, and the realization that this misclassification can be corrected using integral-field stellar kinematics, led to the construction of the \emph{kinematic} morphology-density relation \citep{Cappellari2011b}. This revised relation revealed two key insights:

\begin{description}
    \item [Genuine Spheroidal Galaxies:] Slow rotators are virtually absent in the field or low-density environments. This contrasts with the traditional morphology-density relation, which suggested that the field contained about 10\% of (now known to be misclassified) elliptical galaxies.
    \item [Distribution of Slow Rotators:] The fraction of slow rotators does not vary gradually with environmental density. Instead, massive slow rotators are almost exclusively found at the density peaks of groups and clusters \citep{Cappellari2011b, Cappellari2013apjl, Scott2014, DEugenio2013, Fogarty2014}.
\end{description}

Subsequent analyses from the SAMI and MaNGA surveys initially appeared to challenge earlier findings by reporting \emph{no significant environmental dependence} of early-type galaxy (ETG) kinematics when controlling for stellar mass \citep{Brough2017, Greene2017, Veale2017}. These studies argued that stellar mass alone governs $\lam$ evolution, with environment playing negligible role. However, this interpretation overlooks a critical coupling: galaxy mass growth and angular momentum loss are inherently linked processes driven by mergers. Environmental effects (e.g., cluster vs. field habitats) modulate merger rates, which simultaneously drive (i) stellar mass growth through mergers and (ii) angular momentum decrease via dynamical heating.  

This coupling naturally explains the initially observed insignificant $\lam$--environment trends \emph{at fixed $M_*$} -- galaxies cannot lose angular momentum without gaining mass through the same environmental processes. Furthermore, three limitations compound this null result: (i) slow rotators are intrinsically rare, reducing statistical power; (ii) environmental metrics (e.g., halo mass, local density) carry large systematic uncertainties.  

A biological analogy clarifies this problem: attempting to detect environmental effects on human height \emph{at fixed weight} would be equally challenging. Environment modulates growth processes that simultaneously increase both height (through nutrition) and weight (through calorie intake) during development. Just as childhood nutrition affects adult stature \emph{through} its association with weight gain, galaxy environments shape $\lam$ \emph{through} their role in mass assembly. Expecting strong environmental signatures at fixed $M_*$ is thus both statistically and physically ill-posed -- the two parameters represent different facets of the same evolutionary history. The apparent contradiction between studies instead reinforces the unified $\lam$ and $M_*$ evolutionary sequence predicted by hierarchical models.  

Later, more detailed analysis of both MaNGA and SAMI survey data uncovered the expected weak \emph{residual} trends of ETG kinematics as a function of environment, \emph{at fixed stellar mass}, on top of the well-established major dependency as a function of mass \citep{Graham2019_environment, vandeSande2021_environment}. This is illustrated in \autoref{fig:kinematics_fixed_mass}: the left panel shows the main trend, with the fraction of slow rotator ETGs increasing sharply above the critical mass stellar $\mcrit\approx2\times10^{11}\msun$ \citep{Cappellari2016} and dominating at $M_*\sim10^{12}\msun$. In the figures, each mass bin was chosen to contain the same number of about 200 galaxies, to ensure that number statistics do not impact any observed difference. The middle panel shows how the fraction of slow rotators increases with environmental density \emph{at fixed stellar mass}. Especially in the largest mass bins, where the fraction of slow rotators is larger, there is a clear increase as a function of environment. The right panel shows the mass function of fast/slow rotator ETGs for three environmental densities. It shows that, while the fraction of fast rotators and spirals is independent of environmental density, the mass function of slow rotators changes as a function of environment, shifting towards the largest masses in the densest environments. A similar result was shown for a complete but smaller sample, with different data, in \citet[fig.~4]{Cappellari2013apjl}.

Given the rarity of slow rotators in clusters, a more direct way of understanding the effect of the environment on their distribution can be obtained by directly looking at a small set of well-studied clusters. This is illustrated in \autoref{fig:slow_rotator_in_clusters}, where the slow rotators appear to trace the peaks of the density distribution in clusters. Given that clusters form hierarchically by the merging of smaller groups, the fact that slow rotators lie close to the density of substructure suggests they were already formed in the smaller groups before these merged to form larger clusters. Similar slow rotator ETG distributions were shown with complete kinematic coverage for the well-studied Fornax and Virgo clusters in the review by \citet[fig.~26]{Cappellari2016} and with more sparse kinematic coverage in eight clusters by \citet{Brough2017}.

\section{Formation and Evolution: A Tale of Two Paths}

The bimodal distribution of the photometric and kinematic properties of early-type galaxies (ETGs) reflects the different channels shaping their evolution. Comprehensive reviews of galaxy formation and key challenges are provided by \citet{Somerville2015, Naab2017}, while \citet{Cappellari2016} reviews specifically the evolutionary paths of fast and slow rotator ETGs. To understand the distinct properties of fast and slow rotator ETGs, we must consider the hierarchical growth of galaxies, clusters, and their dark halos \citep{Mo2010}.

In the early universe, primordial density fluctuations led to the formation of dark matter halos, which attracted and accumulated gas. As the gas cooled and condensed, it formed rotating disks that eventually became spiral galaxies \citep{White1978}. The most massive halos, created from the largest density fluctuations, attracted gas at much higher rates due to their strong gravitational pull. This resulted in violent disk instability, nuclear gas inflows, and intense star formation. The resulting supernova explosions from massive stars caused strong feedback on the gas and enhanced the $\alpha$ elements. Additionally, when halos exceeded a critical mass, the infalling gas was shock-heated \citep{Keres2005, Dekel2006}.

At the centers of these massive halos, supermassive black holes grow from pre-existing seeds, accreting gas and producing jets that inject energy into the surrounding medium \citep[e.g.,][]{Dubois2016, Nelson2019}. Initial indications from JWST \citep{Maiolino2024} suggest that by redshift $4\lesssim z\lesssim7$, black holes lie on the local relation between stellar and black hole mass. These combined feedback effects rapidly suppress star formation, leading to the formation of the precursors of slow rotator ETGs, situated at the centers of their massive dark halos and surrounded by hot gas glowing in X-rays \citep[e.g.,][]{Bender1989}. Due to intense and chaotic gas accretion and gas disk disruption, these massive galaxies are characterized by their spheroidal shapes, lack of stellar disks, and low specific angular momentum.

The subsequent evolution of slow rotators follows the hierarchical growth of galaxy groups into larger clusters \citep[e.g.,][]{DeLucia2012}. Because slow rotators quench their star formation early and inhabit environments hostile to further cold gas accretion, their evolution is dominated by gas-poor (dry) mergers. With their large masses and low relative velocities within their halos, slow rotators have large collisional cross-sections. Consequently, when their host groups or clusters merge, they also merge efficiently while also accreting smaller quenched galaxies, forming even more massive slow rotators \citep[e.g.,][]{Choi2018}. These massive slow rotators have extended outer stellar halos and high Sérsic indices (\autoref{fig:core_sersic}) due to the deposition of accreted stars at large radii \citep[e.g.,][]{Naab2009, Pillepich2018}. During gas-free mergers, galaxies are expected to move along lines with $M_*\propto\re$ (or slightly steeper) \citep{Naab2009,Bezanson2009}, without altering their chemical composition since no new star formation is involved. This explains the near-constancy of galaxy properties along lines of constant $\se\propto\sqrt{M_*/\re}$ on the $(M_*,\re)$ diagram (\autoref{fig:mass-size}, \autoref{fig:mass_size_cappellari2016}). The growth in size \citep[e.g.,][]{Trujillo2006gems, vanderWel2014} and Sérsic index of massive galaxies can be traced as a function of redshift \citep[e.g.,][]{vanDokkum2010profiles} \citep[see review in][sec.~6]{Cappellari2016}.

When slow rotators merge, their supermassive black holes sink to the center of the remnant through dynamical friction, forming a black hole binary that eventually merges to produce a larger black hole, while emitting detectable gravitational waves. During this process, the black hole binary ejects stars on radial orbits \citep{Milosavljevic2001, Rantala2024}, leaving a `scoured' flattened inner stellar core (\autoref{fig:core_sersic}) with a tangentially-biased nuclear stellar orbital distribution (\autoref{fig:anisotropy_radial}). This hierarchical growth process, including the two-phase evolution with intense in-situ star formation followed by dry merger accretion \citep{Oser2010}, can accurately reproduce the observed properties of slow rotators \citep{Rantala2024}, including their location at cluster density peaks (\autoref{fig:slow_rotator_in_clusters}). The significant enhancement in $\alpha$ elements \citep{Krajnovic2020} provides a glimpse into the early epoch of highly efficient star formation within these massive halos at high redshift.

In contrast, gas-rich disk galaxies in lower-density environments evolve more gradually \citep[e.g.,][]{Pillepich2019}. These galaxies may undergo gas-rich mergers, triggering bursts of star formation and leading to the formation of stellar bulges as gas dissipates and sinks into their centers. During these mergers, stellar density increases near the centers, the central SMBH grows by accretion, and the stellar populations evolve due to fresh star formation and chemical recycling, increasing their metallicity. This implies that these galaxies will cross lines of constant $\se\propto\sqrt{M_*/\re}$, from left to right (\autoref{fig:mass_size_cappellari2016}; \citealt[fig.~29]{Cappellari2016}). However, as these galaxies evolve, they may become quiescent (quenched) due to internal feedback mechanisms, such as AGN activity and stellar winds from supernova explosions. Empirical evidence of this quenching process out to $z\sim3$ has been detected with JWST \citep{DEugenio2024quench}. These passive galaxies, the fast rotator ETGs, are characterized by their disk-like morphologies (like S0s or disky ellipticals, when seen edge-on) and high specific stellar angular momentum.

The environmental impact on the evolution of galaxies with low and intermediate masses is profound. Disk galaxies are uniformly scattered inside groups and clusters with relative low masses, large relative velocities and consequently small collisional cross section. In dense environments, like galaxy clusters, they are subjected to tidal forces and ram pressure stripping, which can remove their gas and quench star formation \citep[see reviews by][]{Boselli2006, Cortese2021}. Combined with internal processes, these environmental effects transform disk galaxies into fast rotator ETGs, which may subsequently accrete more mass through minor mergers. This is evidenced, for instance, by integral-field spectroscopic observations showing galaxy disks with lower metallicity than their bulges, likely indicating fresh, low-metallicity gas accretion \citep{Lu2023}.

While this simplified picture provides a framework for understanding ETG formation and evolution, the reality is more complex. Various factors, including initial conditions, merger history, and environmental influences, shape these galaxies' final properties. Future observations and simulations will continue to illuminate the intricate processes that govern the birth and evolution of early-type galaxies.

\section{Summary and Outlook}

In this chapter, I have shown that early-type galaxies (ETGs) are not a homogeneous class but display a bimodal distribution in key structural properties, particularly in their stellar specific angular momentum, kinematic morphology, and nuclear surface brightness profiles.

\begin{description}
    \item[Slow Rotator ETGs:] This class of ETGs tends to be found at the peaks of local overdensities in galaxy clusters. There is a strong dependency between stellar mass and specific angular momentum. Slow rotators are nearly absent at low mass but start to dominate above a characteristic stellar mass of $\mcrit\approx2\times10^{11}\msun$. They are generally close to spherical or mildly triaxial within $1\re$ and can become significantly flattened only in their outer stellar halos. These galaxies are metal-rich, composed of stars nearly as old as the Universe, and enhanced in $\alpha$ elements, indicating they formed most of their stars quickly and efficiently at high redshift, growing primarily through gas-poor mergers.
    
    \item[Fast Rotator ETGs:] This class is closely related to spiral galaxies, forming a continuous sequence of properties with them. The structure of fast rotator ETGs cannot be understood without considering spiral galaxies. Empirically, this sequence starts with star-forming spiral galaxies, continues with S0 galaxies, and ends with the most spheroid-dominated disky elliptical galaxies. The empirical global parameter that best traces this sequence is the stellar velocity dispersion $\se$. Equally effective tracers of this sequence from spiral to fast rotator ETGs include the virial estimator of $\se$ (inferred from stellar mass and size using $\se\propto\sqrt{M_*/\re}$) or the surface density $\Sigma_1$ within 1 kpc. There is a critical value of $\lg(\se^\mathrm{crit}/\kms)\approx2.3$, above which all ETGs are quenched, passive, and metal-rich. However, below $\se^\mathrm{crit}$, this parameter does not fully capture the variation in galaxy properties; instead, at smaller $\se$ values, there is a spread in age at fixed $\se$, with a clear trend for younger galaxies to have lower metallicity.
\end{description}

The distribution of fast/slow rotator ETGs and spiral galaxies as a function of their environment, their structural differences, and the modeling of galaxy evolution via numerical simulations indicates two broadly distinct paths for their formation. Slow rotators form most of their stars early and rapidly in the Universe's evolution. They quench quickly and subsequently remain passive for most of their evolution, growing via dry mergers and following the hierarchical growth of their host groups/clusters. Fast rotators start as star-forming disks, growing more gradually over time. Their bulges and supermassive black holes co-evolve, fed by accreted gas, until they are quenched by internal feedback or environmental factors.
 
Our understanding of ETGs' structure and evolution is built on a synergy between detailed local observations, mainly from integral-field spectroscopy, and less detailed observations at higher redshifts, generally involving photometry alone. To test the evolutionary scenario, we need integral-field spectroscopy observations at higher redshifts. Until now, these have been beyond the reach of current instrumentation. However, with the launch of JWST, high-spatial-resolution integral-field spectroscopy is now possible. This allows for spatially-resolved stellar kinematics up to redshift $z\sim2$, at the critical time when most of the mass is being assembled. In a few years, the Extremely Large Telescope will enable even higher spatial resolution integral-field spectroscopy, promising a new revolution in galaxy formation studies.

\renewcommand*{\rmdefault}{\sfdefault}
\printbibliography

\end{document}